\documentclass[namedreferences]{solarphysics}
\usepackage[hyperref,optionalrh,solaromanenum,natbib]{spr-sola-addons} % For Solar Physics 
\usepackage{natbib}

\usepackage{wasysym}

\usepackage{amsmath}
\usepackage{amsfonts}
\usepackage{graphicx}                    % For eps figures, newer & more powerf ull
\usepackage{epsfig}                     % For eps figures, old commands
\usepackage{courier}                    % Change the \texttt command to courier style
\usepackage{amssymb}                    % useful mathematical symbols
\usepackage{flexisym}
\usepackage{color}
\usepackage{epstopdf}
\usepackage{url}                         % For breaking URLs easily trough lines
\usepackage{pdflscape}
\renewcommand{\vec}[1]{{\mathbfit #1}}

\newcommand{\aap}{    {\it Astron. Astrophys.}}
\newcommand{\aaps}{   {\it Astron. Astrophys. Suppl.}}
\newcommand{\aapr}{   {\it Astron. Astrophys. Rev.}}

\newcommand{\apj}{    {\it Astrophys. J.}}
\newcommand{\apjl}{   {\it Astrophys. J. Lett.}}

\newcommand{\solphys}{{\it Solar Phys.}}
 
\newcommand{\ssr}{    {\it Space Sci. Rev.}} 
\newcommand{\apjs}{   {\it Astro Phys. Suppl. Series}} 
\newcommand{\aop}{   {\it Applied Optics}}
\newcommand{\procspie}{   {\it Proceedings of SPIE International Conference on Optics and Photonics 2015}}

\begin{document}
\begin{article}

\begin{opening}

\title{Imaging Spectropolarimeter for Multi-application Solar Telescope at Udaipur Solar Observatory: Characterization of polarimeter and preliminary observations}

\author{Alok Ranjan~\surname{Tiwary}$^{1}
                       $\sep Shibu K.~\surname{Mathew}$^{1}
                       $\sep A. Raja~\surname{Bayanna}$^{1}
                       $\sep P.~\surname{Venkatakrishnan}$^{1}
                       $\sep Rahul ~\surname{Yadav}$^{1}$
        }

\runningauthor{Tiwary et al.}
\runningtitle{Polarimeter for MAST}

\institute{$^{1}$ Udaipur Solar Observatory, Physical Research Laboratory, Udaipur, Rajasthan, India, 313001 \\
                     email: \url{atiwary@prl.res.in} 
             }

\begin{abstract}

\textit{Multi-Application Solar Telescope} (MAST) is a 50 cm off-axis Gregorian telescope and started operational recently at \textit{Udaipur Solar Observatory} 
(USO). For understanding the evolution and dynamics of solar magnetic and velocity fields, an imaging spectropolarimeter is being developed as one of the back-end instruments of MAST. This system consists 
of a narrow-band filter and a polarimeter. Polarimeter includes a linear polarizer and two sets of Liquid Crystal Variable Retarders (LCVRs). 
The instrument is intended for the simultaneous observations in the spectral lines 6173 \AA{} and 8542 \AA{}, which are formed in photosphere and chromosphere, 
respectively. In this paper, we present results from the characterization of the LCVRs for the spectral lines of interest and response matrix of the polarimeter. 
We also present preliminary observations of an active region obtained using the spectropolarimeter. For verification, we compare the Stokes observations of the 
active region obtained from \textit{Helioseismic Magnetic Imager} (HMI) onboard \textit{Solar Dynamics Observatory}(SDO) with that of MAST observations in 
the spectral line 6173 \AA{}. We found good agreement between both the observations, considering the fact that MAST observations are seeing limited.

\end{abstract} 
\end{opening}
\section{Introduction}
Solar activity is driven by spatio-temporal distribution of magnetic field (e.g. \cite{Solanki2003}, \cite{Borrero2011}, \cite{Stix2004}). Therefore, the precise 
measurements of the magnetic field in the solar atmosphere is of fundamental importance. The Zeeman effect \citep{Zeeman1897} has been recognized as the most 
authentic tool for more than a century to derive the magnetic field of sunspots and pores in the photosphere \citep[and references therein]{stenflo2015}. With the 
advancement of instrumentation capabilities, one can now measure the Zeeman signals of a small-scale structure present in the photosphere as well as in the 
chromosphere \citep{2014ASPC..489..137T, Wiegelmann2015}. However, for the estimation of the weak field in the corona, the Zeeman 
effect becomes completely insensitive. Therefore, we rely on the Hanle effect \citep{Hanle1924} for coronal magnetic field measurements. In special cases, both 
these effect can be combined to have magnetic field measurements in different atmospheric layers of the Sun \citep{Lagg2015}. \newline

In the presence of magnetic field, light gets polarized and depolarized. 
The polarization of light is described in terms of the Stokes parameters I, Q, U, and V \citep{1960ratr.book.....C, ob:bornwolf} where I give the total intensity,
Q and U represent the linear polarization and V represents the circular polarization. 
The spectropolarimeter consists of a filtergraph/spectrograph and a polarimeter, is employed to derive the solar vector magnetic fields by measuring Stokes
I, Q, U, and V. Generally, two different techniques have been commonly used for spectral analysis: (1) imaging (filter) based (2) slit-based. For polarization analysis either single beam polarimeter or dual-beam polarimeter is employed.
In imaging based spectropolarimetry, 2-D images are obtained in a sequence by tuning a narrow-band filter 
to different wavelengths along the spectral line profile of interest. Modern imaging spectropolarimeters employ either single or multiple Fabry-Perot (FP) etalon as narrow-band 
filters because of their high transmission and fast tuning capability. 
Few examples of the currently working imaging spectropolarimeters are: GREGOR Fabry-Perot instrument (GFPI; 
\cite{2010SPIE.7735E..6MD}), KIS/IAA Visible Imaging Polarimeter (VIP; \cite{Beck2010a}), CRisp Imaging 
Spectro-Polarimeter (CRISP, \cite{Scharmer2008}), the Gottingen spectropolarimeter \citep{BelloGonzalez2008}, 
Interferometric BIdimensional Spectrometer (IBIS; \cite{Cavallini2006}) and  Imaging Vector Magnetograph 
(IVM; \cite{Mickey1996}). The Imaging Magnetograph eXperiment (IMaX; \cite{MartinezPillet2011b}) flown with the  
the Sunrise balloon mission \citep{Barthol2011} is another example of imaging spectropolarimeter using a voltage tunable Lithium Niobate Fabry-Perot etalon.
These instruments differ in the number of etalons and the optical configuration 
(telecentric or collimated). On the other hand, slit-based spectropolarimeter obtains the spectrum 
by scanning the required field-of-view (FOV) in sequence. Examples of slit-based spectropolarimeters are\textbf{:} Diffraction Limited Spectro-Polarimeter 
(DLSP; \cite{2003SPIE.4843..414S}), the POlarimetric LIttrow Spectrograph (POLIS; \cite{Beck2005}), the Spectro-Polarimeter for Infrared and Optical Regions 
(SPINOR; \cite{Socas-Navarro2006}), and spectropolarimeter of the \textit{Solar Optical Telescope} (SOT) onboard Hinode (\cite{Ichimoto2008}, \cite{Tsuneta2008}). 
Based on the scientific requirement either of the above mentioned techniques are preferred. 
However, with the advances in the technology, both these techniques yield similar results.
\newline
In order to measure the vector magnetic field in the solar atmosphere, we have developed an imaging spectropolarimeter 
for \textit{Multi Application Solar Telescope} (MAST) \citep{2008SPIE.7012E..35D, 2010SPIE.7733E..35D, Mathew2009} which is recently installed at the lake site of 
\textit{Udaipur Solar Observatory} (USO). It is an off-axis Gregorian telescope with a 50 cm aperture. Along with adaptive optics system, the telescope is designed 
to provide near diffraction limited observations. One of the scientific objectives of MAST is to study the evolution of vector magnetic field in the solar atmosphere at different heights and its connection to various solar activities. Imaging spectropolarimeter for MAST consists of a narrow-band imager \citep{RajaBayanna2014} and a polarimeter which are used to measure the Stokes vector at two 
different wavelengths \textit{i.e.} at 6173 \AA{} and 8542 \AA{}, corresponding to photospheric and chromospheric heights, respectively.
\newline
This paper has been arranged in the following manner. Section 2 describes the design of the polarimeter and the modulation scheme to measure the Stokes vector. 
The characterization of the liquid crystal variable retarders (LCVRs) with voltage and temperatures are discussed in Section 3. In Section 4 we explain the response 
matrix of the polarimeter derived using an experimental setup in the laboratory. Preliminary observations obtained with our instrument are presented in Section 5.
Summary of the paper is provided in Section 6.

\section{Polarimeter schematic and components}

A polarimeter measures the polarization of the light by modulating the input polarization into measurable intensities. In general, the polarization analysis 
can be done in two ways \cite[and references therein]{2003isp..book.....D}: 
(a) Temporal polarization modulation/single beam polarimetry: Here, the different polarization measurements are obtained sequentially. The time gap between
the measurements could introduce seeing related spurious signals in the difference image (\cite{Lites1987}; \cite{Leka2001}). This can be minimized either by 
compensating the atmospheric turbulence by adaptive optics or by implementing a very fast modulation scheme, wherein one modulation cycle is completed before 
atmospheric seeing changes completely, or by both. However, this imposes stringent requirements of the polarization modulator. 
(b) Spatial polarization modulation/Dual beam polarimetry \citep{Lites1987}: Here the orthogonal polarization states are separated by means of the polarizing beam 
splitter or displacer and both the beams are recorded simultaneously. This cancels out the fluctuations in the Stokes I to the other Stokes parameters caused due to 
atmospheric seeing \citep{MartinezPillet2011b}. 
Since both the beams are used for final computation of Stokes parameters, this method improves signal-noise-ratio (SNR) by a factor of $\sqrt{2}$ as compared to the single 
beam polarimetry. However, different optical paths for the measurements of two polarization states might introduce a systematic error; this puts a stringent 
requirement on the quality of the two optical paths in the experimental setup. It also requires a larger area of the detector to accommodate larger FOV.

Though the dual beam polarimetry is advantageous than the single beam polarimetry, we preferred to use single beam setup to perform the 
polarization analysis over a larger FOV. The fast modulation scheme with Liquid crystal variable retarders along with a matching fast 
camera readout, enable us to complete the modulation cycle before the seeing changes significantly. With suitable large format camera and a polarizing beam 
displacer, we plan to implement the dual beam spectropolarimetry with MAST at a later stage.

The main components of the MAST polarimeter are LCVRs for polarization modulation and Glan-Thompson polarizer as analyzer.
Many of the recent polarization modulators also use liquid crystals, in which retardance (as in nematic liquid crystals) or fast axis (as in ferroelectric 
liquid crystals) can be changed by applying voltages \citep[and references therein]{Heredero2007}. These modulators allow us to implement fast modulation schemes 
by avoiding mechanical motions and beam wobble 
as in the case of rotating retarders \citep{Heredero2007}. In MAST two sets of LCVRs along with a linear polarizer are used for obtaining the Stokes parameters. 
LCVRs for MAST polarimeter are custom made nematic liquid crystal devices with an aperture size of 80 mm, procured from Meadowlark Optics, USA. Figure 1 shows the 
schematic of MAST polarimeter. The fast axis of LCVR1 and LCVR2 are fixed at $0^{\circ}$ and $45^{\circ}$ with respect to the linear polarizer (LP), respectively. 
Light from the telescope enters through the LCVR1, passes through the LCVR2 and exits from the LP. Both LCVRs and the LP are mounted in rotating mounts to make the 
adjustment for the angles. The photograph of the installed system is shown in Figure 2.  The temperature of the LCVRs is actively controlled using flexible heaters 
fixed on the holder; a temperature sensor in closed loop provides a thermal stability of $\pm$1$^{\circ}C$.

 \begin{figure}    
%    \centerline{\includegraphics[width=0.35\textwidth,clip=]{Figure11.eps}
\centerline{\includegraphics[scale=0.40]{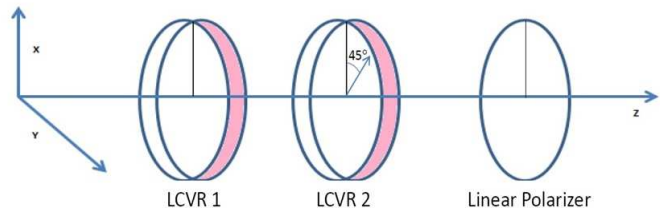}
}
%               \vspace{-0.5\textwidth}
              \caption{Schematic layout of the polarimeter for MAST. The fast axis of LCVR1 and LCVR2 are kept at $0^{\circ}$ 
                            and $45^{\circ}$ with respect to the linear polarizer.}
   %\label{fig1}
   \end{figure}
   
\begin{figure}  
   \centerline{\includegraphics[width=0.65\textwidth,clip=]{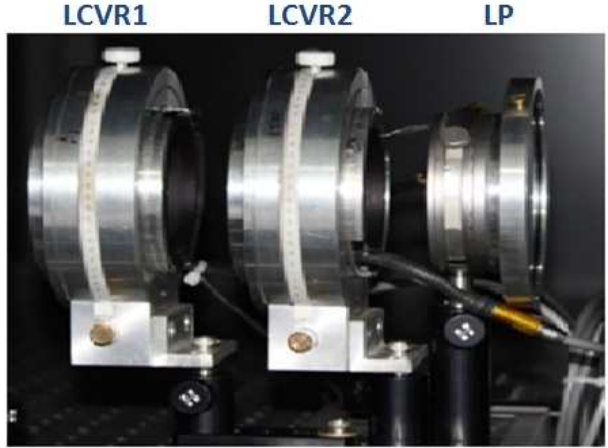}
              }
              \caption{Imaging polarimeter for MAST at USO. From left to right LCVR1, LCVR2 and linear polarizer. The LCVR1 and LCVR2 are kept at 
              $0^{\circ}$ and $45^{\circ}$ with respect to the linear polarizer (LP), respectively.}
   %\label{fig2}
   \end{figure}
The retardance of the LCVR can be changed by applying voltages. The modulation voltages for the LCVRs are supplied from a Meadowlark digital interface. 
The amplitude of the basic 2 KHz square waveform can be adjusted by an input DC voltage or counts provided from the software through a USB computer
interface. Modulation voltages in the range of $0-10$ V with 16-bits accuracy can be applied to the LCVRs from this interface. C-programs 
have been written for the image acquisition in synchronization with the modulation voltages. 
The modulation scheme employed here is described in the following subsection.

\subsection{Modulation scheme for the MAST polarimeter}
As discussed in the previous section, the MAST polarimeter consists of two LCVRs and a linear polarizer. The Stokes vector of input light 
($S_{in}$) at LCVR1 and the Stokes vector of output light at LP ($S_{out}$), are related using Mueller-matrix formalism by the following equation: 
\begin{equation}%\label{eqn1}
S_{out}=M_{P}M_{LC2}(\gamma,\theta_2)M_{LC1}(\delta,\theta_1)S_{in}=M_SS_{in}
\end{equation}
where $M_{P}$, $M_{LC1}(\delta,\theta_1)$, and $M_{LC2}(\gamma,\theta_2)$ are the Mueller matrices of the linear polarizer, LCVR1 and LCVR2, respectively.
 The fast axis of LCVR1 and LCVR2 with respect to the transmission axis of the linear polarizer is represented by $\theta_1$ and $\theta_2$, 
 respectively. The retardance of the LCVR1 and LCVR2 are given by $\delta$ and $\gamma$, respectively. $M_S$ is the resultant Mueller matrix of the polarimeter
 for a particular value of $\delta, \gamma, \theta_1$, and $\theta_2$.

Using the Mueller matrix of the retarder and polarizer in the Equation (1), intensity I, measured at CCD, can be expressed as a linear combination 
of all the input Stokes parameters as shown below.
\begin{equation}%\label{eqn2}
I_j=S_{out}^0=1S_{in}^0+a_jS_{in}^1+b_jS_{in}^2+c_jS_{in}^3,
\end{equation}
where the parameters $[1,a_j,b_j,c_j]$ are the first row of $M_S$ which depend on $\theta_1$, $\theta_2$, $\delta$, and $\gamma$. Here \textit{j} runs over the 
number of observations (from 1 to \textit{n}). From Equation (2), we can measure \textit{n} number of intensities for different combinations of retardances 
and fast axis orientations of both the LCVRs. 
However, it is always preferred to have a minimum number of intensity measurements to infer all the Stokes parameters \citep{delToroIniesta2000}. For vector 
polarimetry a minimum four measurements (\textit{n} $\geq$ 4) are needed whereas longitudinal polarimetry can be done with two measurements (\textit{n} $\geq$ 2) 
only. Therefore for \textit{n} intensity measurements, a modulation scheme is fully characterized by a $(\textit{n}\times4)$ modulation matrix \textbf{O} built 
from the n first rows of $M_S$,
\begin{equation}%\label{eqn3}
I_{meas}=\textbf{O}S_{in} .
\end{equation}
The Stokes vector is derived using the following Equation,

\begin{equation}%\label{eqn4}
\textbf S_{in}=\textbf{O}^{-1}\textbf I_{meas}=\textbf{D} I_{meas} . 
\end{equation}
If \textit{n}$=4$, then \textbf{O} is a $4\times4$ matrix so its inverse will be unique. But, if \textit{n} $=$ 6 then \textbf{O} is $6\times4$ matrix, which is 
not a square matrix, and its inverse will not be unique. Hence, when D is not a square matrix, it is determined using the following Equation \citep{delToroIniesta2000}, 
 \begin{equation}%\label{eqn5}
 \textbf{D}=(\textbf{O}^T\textbf{O})^{-1}\textbf{O}^T .
 \end{equation}
 The efficiency of the modulation scheme is defined as, 
 \begin{equation}%\label{eqn6}
 e_i=(\textit{n}\sum_{j=1}^{\textit{n}}\textbf{D}_{ij}^2)^{-1/2},
 \end{equation}
where, $\textit{e}_1$, $\textit{e}_2$, $\textit{e}_3$, and $\textit{e}_4$ are the efficiencies for measuring the Stokes parameter I, Q, U, and V respectively.
\newline
For four measurement modulation scheme (\textit{n}=4),
\begin{equation}%\label{eqn7}
I_{meas}=
\left[
\begin{array}{c}
I_1\\ I_2\\I_3\\ I_4
\end{array}
\right] 
=
\textbf{O}\left[
\begin{array}{c}
S_{in}^0\\S_{in}^1\\S_{in}^2\\S_{in}^3
\end{array}
\right] ,
\end{equation}
where \textbf{O} is the modulation matrix given as,
\begin{equation}%\label{eqn8}
\textbf{O}
=
\begin{pmatrix}
1 & a_1 & b_1 & c_1\\ 
1 & a_2 & b_2 & c_2\\
1 & a_3 & b_3 & c_3\\
1 & a_4 & b_4 & c_4
\end{pmatrix}.
\end{equation}
\newline
For the polarimeter configuration shown in Figure 1 orientation of the fast axis of LCVR1 and LCVR2 is fixed at $\theta_1=0^{\circ}$, and $\theta_2=45^{\circ}$, 
respectively \citep{2004SPIE.5487.1152M}. Using different combinations of retardances of LCVR1 and LCVR2, \textit{i.e.}, $\delta$ and 
$\gamma$, four and six measurement modulation schemes are implemented for vector polarimetry at MAST. 
The modulation schemes of four and six measurements are listed in Table 1 and 2.

\begin{center}
\begin{table}
% 	\centering
\caption{Four measurement modulation scheme \citep{2004SPIE.5487.1152M} for vector polarimetry.}
	%\label{tab1}
	\begin{tabular}{ccc}
		\hline
		 $\delta$ & $\gamma$& Measured Intensity\\ 
		 ($^{\circ}$) &($^{\circ}$)& $I_{meas}$\\
		 \hline
	 315.0 & 305.264 & $I_1$$=$$I+Q/\sqrt{3}+U/\sqrt{3}+V/\sqrt{3}$ \\   
         315.0 & 54.736  & $I_2$$=$$I+Q/\sqrt{3}-U/\sqrt{3}-V/\sqrt{3}$\\ 
	225.0 & 125.264 & $I_3$$=$$I-Q/\sqrt{3}-U/\sqrt{3}+V/\sqrt{3}$ \\ 
	 225.0 & 234.736 & $I_4$$=$$I-Q/\sqrt{3}+U/\sqrt{3}-V/\sqrt{3}$ \\
 \hline
\end{tabular}
\end{table}
\end{center}

\begin{center}
\begin{table}
	\caption{Six measurement modulation scheme \citep{Tomczyk2010} for vector polarimetry.}
%\label{tab2}
	\begin{tabular}{lll}
		\hline
	 $\delta$ & $\gamma$& Measured Intensity\\ 
	($^{\circ}$) &($^{\circ}$)& $I_{meas}$\\
		 \hline
		180.0  & 360.0 & $I_1=I+Q$\\ 
		180.0  & 180.0 & $I_2=I-Q$\\ 
	        090.0  & 090.0 & $I_3=I+U$\\
	        090.0  & 270.0 & $I_4=I-U$\\
	        180.0  & 090.0 & $I_5=I+V$\\
	        180.0  & 270.0 & $I_6=I-V$\\
		\hline
	\end{tabular}
        \end{table}
\end{center}

The modulation matrix (\textbf{O}) for the four measurement modulation scheme is given by,
\begin{equation}%\label{eqn9}
\textbf{O}
=
\begin{pmatrix}
1 & $0.57735$ & $0.57735$ & $0.57735$\\ 
1 & $0.57735$ & $-0.57735$ &$-0.57735$\\
1 & $-0.57735$ & $-0.57735$ & 0.57735\\
1 & $-0.57735$ & $0.57735$ & $-0.57735$
\end{pmatrix} .
\end{equation}
\newline
For this modulation scheme, the maximum efficiencies are determined from Equations (5) and (6) as $\textit{e}_1=1$, $\textit{e}_2=0.57735$, 
$\textit{e}_3=0.57735$, and $\textit{e}_4=0.57735$.  
\newline
Similarly, the modulation matrix for six measurement modulation scheme can be expressed as a $6\times4$ matrix,
\begin{equation}%\label{eqn10}
\textbf{O}
=
\begin{pmatrix}
1 & 1 & 0 & 0\\ 
1 & $-1$ & 0 & 0\\
1 & 0 & 1 & 0\\
1 & 0 & $-1$ & 0\\
1 & 0 & 0 & 1\\
1 & 0 & 0 & $-1$\\
\end{pmatrix} .
\end{equation}
\newline
For this modulation scheme, the maximum efficiencies are determined as $\textit{e}_1=1$, $\textit{e}_2=0.57735$, $\textit{e}_3=0.57735$, and 
$\textit{e}_4=0.57735$. These are the maximum efficiencies that an ideal polarimeter system can have \citep{delToroIniesta2000}.
\newline
Both the modulation schemes provide equal modulation efficiencies for the measurement of Q, U, and V \citep{DelToroIniesta2012}. 
 Either of these modulation schemes can be used for the measurement of all the Stokes parameters in vector polarimetry. Measurement of all the Stokes parameters 
 is required to obtain the vector magnetic field, whereas the longitudinal magnetic field can be obtained from Stokes I and V only. 
The longitudinal measurement could be done by two intensity measurements. The modulation scheme for longitudinal polarimetry is listed in Table 3.
 \begin{center}
\begin{table}
% \centering
	\caption{Modulation scheme for longitudinal polarimetry}
	%\label{tab3}
	\begin{tabular}{lll}
	\hline
	 $\delta$ & $\gamma$& Measured Intensity\\ 
	($^{\circ}$) &($^{\circ}$)& $I_{meas}$\\
         \hline
	   360.0 & 90.0 & $I_1=I-V$\\   
	   360.0 & 270.0 & $I_2=I+V$\\ 
	 \hline
	\end{tabular}
\end{table}
\end{center}
The relation between the incoming Stokes vector and measured Stokes vector is given as \citep{Beck2005},
\begin{equation}%\label{eqn11}
S_{meas}=\textbf{X}.S_{in},
\end{equation}
where \textbf{X} is the $4\times4$ square matrix known as response matrix which includes all the processes for polarimetric measurement such as
properties of optical components, modulation schemes and their demodulation \citep{deJuanOvelar2014, Beck2005}. Furthermore, \textbf{X} can be expressed as
\begin{equation}%\label{eqn12}
 \textbf{X}=\textbf{D}\textbf{O}.
\end{equation}
\newline
In the response matrix of an ideal polarimeter, the diagonal elements will be unity and the off-diagonal elements will be zero. However, in practice 
off-diagonal element of the response matrix will have non-zero values representing the cross-talk between the Stokes parameters introduced due to several 
reasons \citep{deJuanOvelar2014}.

\section{Characterization of the LCVRs} The retardance of the LCVRs can be tuned by applying voltages. The voltage dependence of retardance of LCVR is given
 by the following Equation \citep{Saleh2007},
\begin{equation}%\label{eqn13}
 \delta=\frac{2\pi d}{\lambda}\left(\left(\frac{sin^2(\Theta)}{n_o^2}+\frac{cos^2(\Theta)}{n_e^2}\right)^{-\frac{1}{2}}-n_e\right), 
\end{equation}
where, $n_o$ and $n_e$ are the refractive indices of the ordinary and extraordinary beam and the equilibrium tilt angle $\Theta$ for liquid crystal molecules 
depends on the applied voltage, which is described by
\begin{align*}
\Theta&= 0,\hspace{0.2cm} \forall \hspace{0.2cm} V\leq V_C\\
       &=\frac{\pi}{2}-2\tan^{-1}\left(\exp\left(-\frac{V-V_C}{V_0}\right)\right),\hspace{0.2cm} \forall \hspace{0.2cm} V > V_C ,
\end{align*}
where, $V$ is the applied voltage, $V_C$ is the critical voltage at which the tilting process begins, and $V_0$ is a constant. 
\newline
The change in the retardance of LCVR due to temperature can be expressed by a semi-empirical formula \citep{Li2004, Capobianco2008},
\begin{equation}%\label{eqn14}
   \delta=\delta_0\,A=\delta_0\left(1-\frac{T}{T_C}\right)^{\alpha}
\end{equation}
where $\delta_0$ is the retardance for A=1; A denotes an order parameter which describes the orientational order of nematic liquid crystal. 
For completely random and isotropic liquid crystals A=0, whereas for a perfectly aligned sample A=1 \citep{Haller1975}. For a typical liquid crystal sample, 
A is in the range of 0.3 to 0.8 and generally decreases with temperature \citep{1975JMoSt..29..190J, Shin-TsonWu2001}.
% \footnote{$https://en.wikipedia.org/wiki/Liquid crystal$}. 
In particular, a sharp drop of A to zero value is observed when the system undergoes a phase transition from the liquid crystal phase into the isotropic phase. Thus, $T_C$ is the 
nematic-isotropic transition temperature and $\alpha$ is critical exponent related to the phase transition.
In order to use the LCVRs in the modulation scheme discussed in the previous Section, it is important to have the exact knowledge of retardance and 
its dependence on voltage and temperature.

\subsection{Experimental setup and theory}
Experimental setup for the characterization of LCVRs is shown in Figure 3. We have used a stabilized DC 
lamp as a white light source (S) along with a diffuser to get uniform intensity. A lens (C1)  is placed 
in front of the pinhole to collimate the light. An interference filter (F1) is used  to select the particular wavelength for which 
characterization of the LCVRs is carried out. The LCVR is placed in between  two Glan-Thompson polarizing prisms (P1 \& P2). 
Glan-Thompson polarizing prisms offers high extinction ratio (1$\times$$10^{-6}$) and high transmittance 
for a wavelength  band  between 3500-23000 \AA{} \citep{2001opt4.book.....H}. Another lens (C2) is placed after the 
analyzer (P2) to image the beam onto a CCD camera to measure the intensity of output light. The incoming light is linearly 
polarized by the polarizer P1.  After being retarded by the LCVR,  whose fast axis is at 45$^{\circ}$ the light is analyzed by 
the second polarizer P2. Polarizer P2 is mounted on a computer  controlled rotation stage to measure the intensities at two 
different orientations. 
\begin{figure}  
\centerline{\includegraphics[width=0.9\textwidth,clip=]{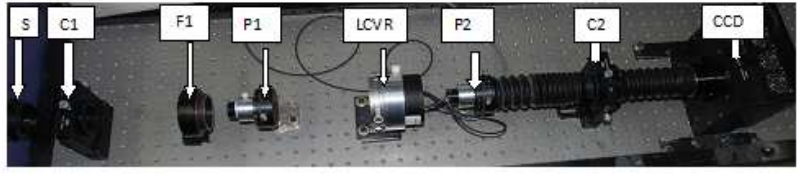}
              }
\caption{Experimental setup for the calibration of LCVRs: In this setup, a light beam coming from a pinhole source is collimated using a lens C1. 
Then the collimating light passes through an interference filter (F1), linear polarizer (P1) and LCVR whose fast axis is kept at $45^{\circ}$ with respect to P1. 
Finally, light is imaged by imaging lens (C2) on CCD which is placed in the focal plane of the imaging lens after passing through analyzer P2.}
%\label{fig3}
\end{figure}
The Stokes vector \textbf{$S_{out}$} after P2 can be written as
\begin{equation}%\label{eqn15}
\textbf{$S_{out}$}=M_{POL}M_R(\delta)\textbf{$S_{in}$},
\end{equation}
where $S_{in}=[1 1 0 0]^T$ represents the light linearly polarized by the polarizer P1. $M_R(\delta)$ and $M_{POL}$ are the Mueller matrix for the LCVR and
 the polarizer P2, respectively.  $M_R(\delta)$ and $M_{POL}$ can be written as
\begin{equation}%\label{eqn16}
 M_{R}(\delta)=
\begin{pmatrix}
1&0&0&0\\
0&cos^{2}2\phi+sin^22\phi cos\delta&sin2\phi cos2\phi(1-cos\delta)&-sin2\phi sin\delta \\
0&sin2\phi cos2\phi(1-cos\delta) & sin^2 2\phi+cos^2 2\phi cos\delta & cos2\phi sin\delta\\
0 & sin2\phi sin\delta & -cos2\phi sin\delta & cos\delta
\end{pmatrix}.
\end{equation}
\begin{equation}%\label{eqn17}
M_{POL}(\theta) =\frac{1}{2} 
\begin{pmatrix}
1&cos2\theta&sin2\theta&0\\ 
cos2\theta&cos^{2}2\theta& sin2\theta cos2\theta&0\\
sin2\theta & sin2\theta cos2\theta & sin^2{2}\theta &0\\
0&0&0&0
\end{pmatrix}.
\end{equation}
where, $\phi$ is the orientation of the fast axis with respect to P1, $\delta$ is the retardance of LCVR due to the voltage applied, and $\theta$ is the orientation 
angle of P2 with respect to P1.
\newline
Substituting $M_R$, $M_{POL}$ and  $S_{in}$ in Equation (15), the output intensity ($I_{out}$) measured by the CCD is 
\begin{equation}%\label{eqn18}
 I_{out}=\frac{1}{2}\left[1+cos2\theta(cos^2 2\phi+sin^2 2\phi cos\delta)+sin2\theta sin2\phi cos2 \phi(1-cos\delta)\right] .
\end{equation}
 In our setup, we always keep $\phi$=$45^\circ$. Hence, Equation (18) becomes
 \begin{equation}%\label{eqn19}
 I_{out}=\frac{1}{2}\left[1+cos2\theta cos\delta \right] .
 \end{equation}
For $\theta= 0^\circ$, where P2 is parallel to P1, output intensity is
\begin{equation}%\label{eqn20}
I_{out}^0=\frac{1}{2}\left[1+cos\delta \right].
  \end{equation}  
For $\theta =90^\circ$, where P2 is crossed to P1, the output intensity is
  \begin{equation}%\label{eqn21}
  I_{out}^{90}=\frac{1}{2}\left[1-cos\delta \right].
  \end{equation}  
Using Equations (20) and (21), retardance of LCVR can be obtained as
  \begin{equation}%\label{eqn22}
  \delta= cos^{-1}\left( \frac{I_{out}^0-I_{out}^{90}}{I_{out}^0+I_{out}^{90}}\right).
  \end{equation}
  \newline
Thus, by measuring the intensities $I_{out}$ for two different angles, $\theta=90^{\circ}$ and $\theta=0^{\circ}$, by rotating the 
polarizer P2, retardance $\delta$ of the LCVR can be estimated for different voltages.  However, it requires prior knowledge of transmission 
axis of polarizers and the fast axis of LCVR. Hence, we determined the  transmission axis of polarizers and the fast axis of LCVR using the 
same experimental setup before we carried out the characterization of LCVR.

\begin{figure}  
   \centerline{\includegraphics[width=0.5\textwidth,clip=]{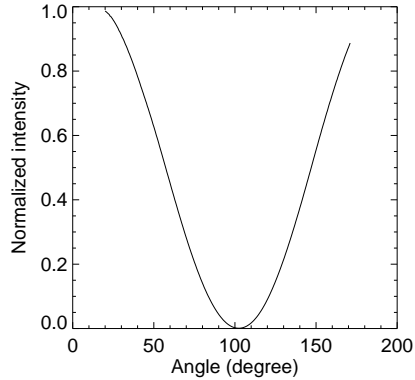}
              }
              \caption{Variation of intensity with the rotation of polarizer P2 with respect to polarizer P1. The minimum intensity gives the
              angle at which both the polarizers are crossed to each other.}
   %\label{fig4}
   \end{figure}
\subsection{Determination of the crossed position of polarizers}
In order to precisely align the axis of the polarizers P1 and P2, LCVR, shown in Figure 3, is 
removed from the optical path. By keeping polarizer P1 at a fixed position, polarizer P2 is rotated 
from $0^{\circ}$ to $180^{\circ}$ with respect to P1, with a step size of 1$^{\circ}$ using a computer controlled 
rotation stage. For each angle, the  intensity is measured using the CCD. The plot between the angle 
and the measured intensity is shown in Figure 4. Initially, when P1 and P2 are in parallel position 
%($10^{\circ}$), 
the intensity is maximum. The intensity starts decreasing with the increase in the angle between them. At crossed position %($100^{\circ}$) 
we get the minimum intensity which further starts increasing with the increase in the angle between P1 and P2.  
The above result satisfies the following equation 
$$
M_{P1}(\theta=0^{\circ}).M_{P2}(\theta=90^{\circ})=0.
$$

\subsection{Determination of the fast axis of LCVR} Knowing the crossed position of polarizers P1 and P2, we proceed to determine the fast axis of 
 LCVR. For this purpose, polarizers P1 and P2 are kept in crossed position and LCVR is again placed between them. After that, output intensity 
 is measured by rotating the LCVR. Figure 5 shows the plot between measured intensity and angle of the LCVR with respect to P1.
 The angle at which minimum intensity is observed is the angle at which fast axis of LCVR is parallel to P1. 
 By knowing this angle, we rotate the fast axis of LCVR to 45$^{\circ}$ with respect to P1 for further characterization of the LCVR.
 The above procedure can be easily understood by solving the following Mueller matrix Equation
$$
M_{P1}(\theta=0^{\circ}).M_{LCVR}(\phi=0^{\circ},\delta).M_{P2}(\theta=90^{\circ})=0 .
$$

\begin{figure}
        \centering
	\includegraphics[width=0.60\textwidth]{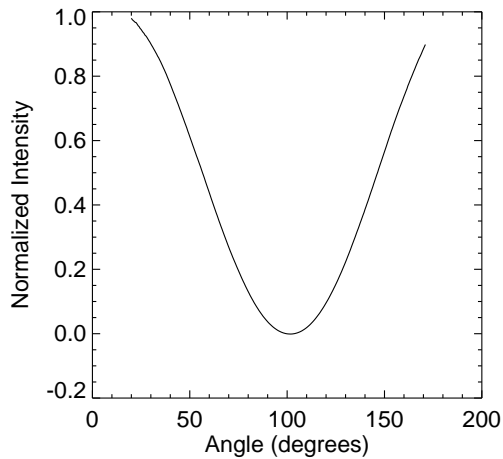}
	\caption{Variation of intensity with the rotation of the fast axis of LCVR. The minimum intensity gives the position where the fast axis
	of LCVR is parallel with the polarizer P1.}
	%\label{fig5}
\end{figure}

  \begin{figure}    
   \centerline{\hspace*{0.015\textwidth}
               \includegraphics[width=0.515\textwidth,clip=]{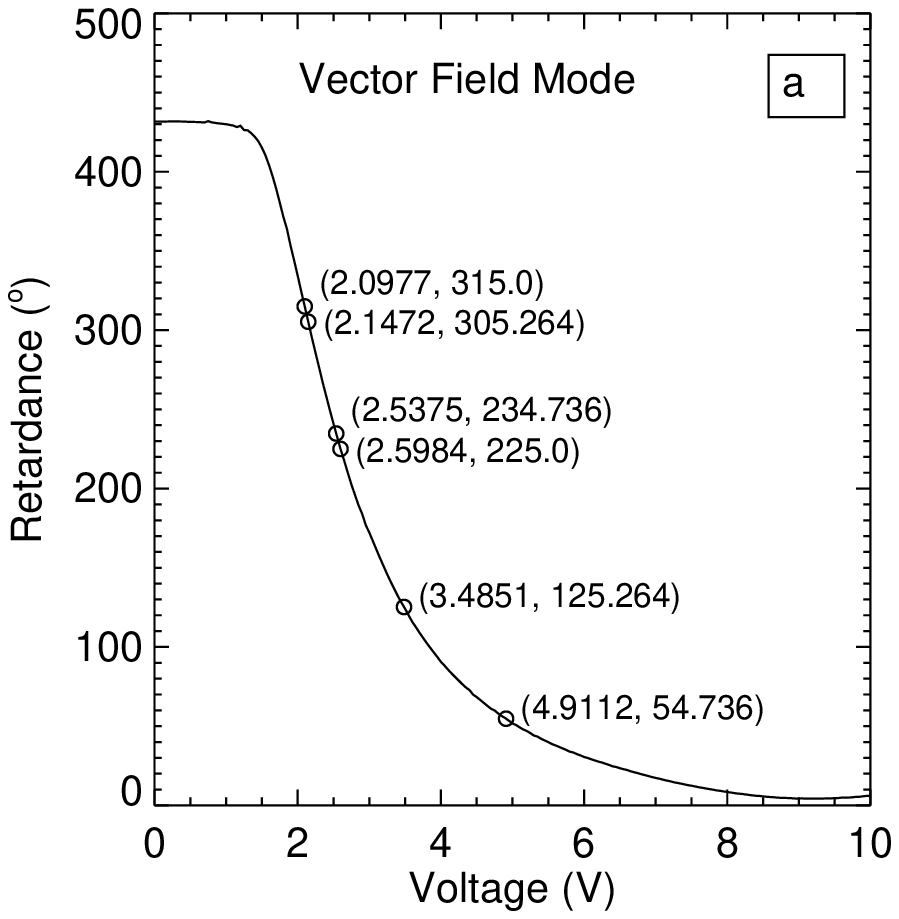}
               \hspace*{-0.03\textwidth}
               \includegraphics[width=0.515\textwidth,clip=]{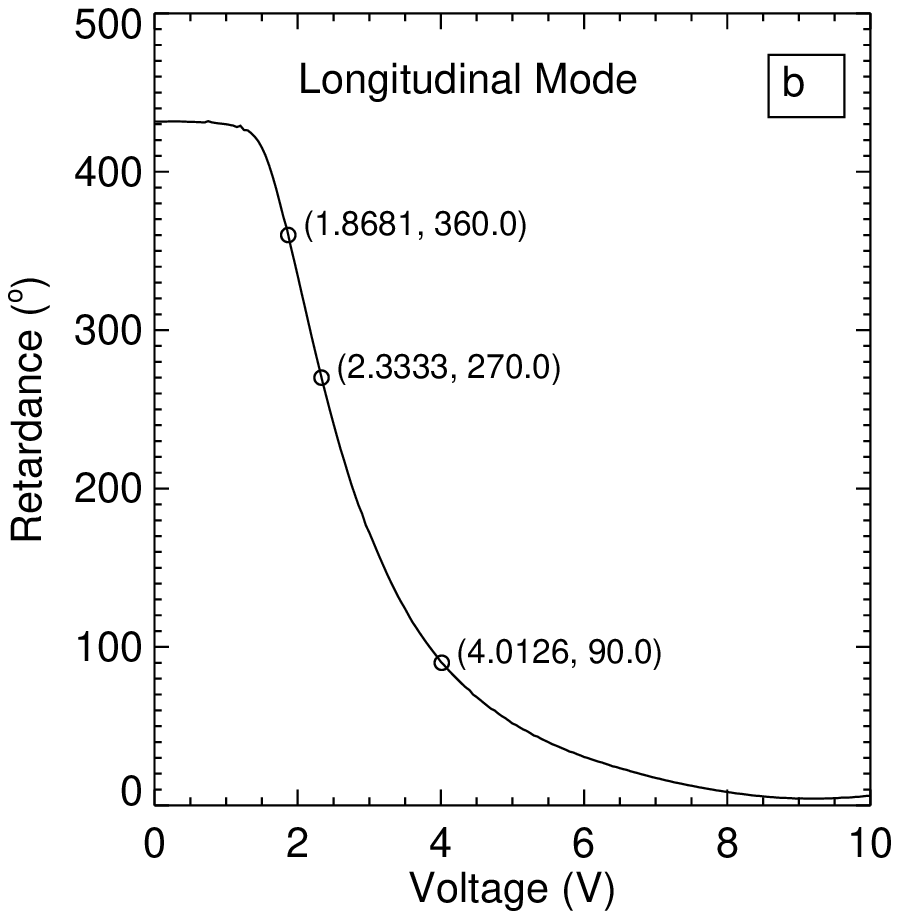}
              }
     \vspace{-0.35\textwidth}   % Shift close to the panel top 
     \centerline{\Large \bf     % Includes the labels (here needs the color 
                                %   package, see beginning of this file)
      \hspace{0.0 \textwidth}  %\color{white}{}
      \hspace{0.415\textwidth}  %\color{white}{}
         \hfill}
     \vspace{0.31\textwidth}    % Shift back to the panel bottom 
          
  \centerline{\hspace*{0.015\textwidth}
             \includegraphics[width=0.515\textwidth,clip=]{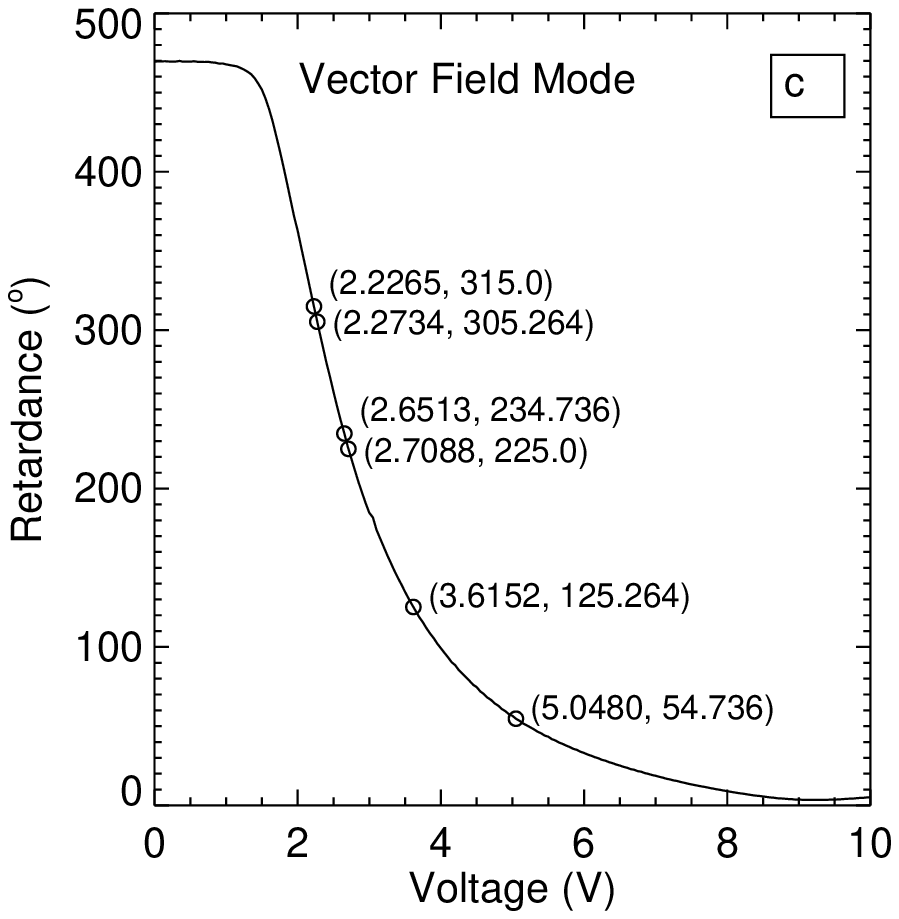}
             \hspace*{-0.03\textwidth}
              \includegraphics[width=0.515\textwidth,clip=]{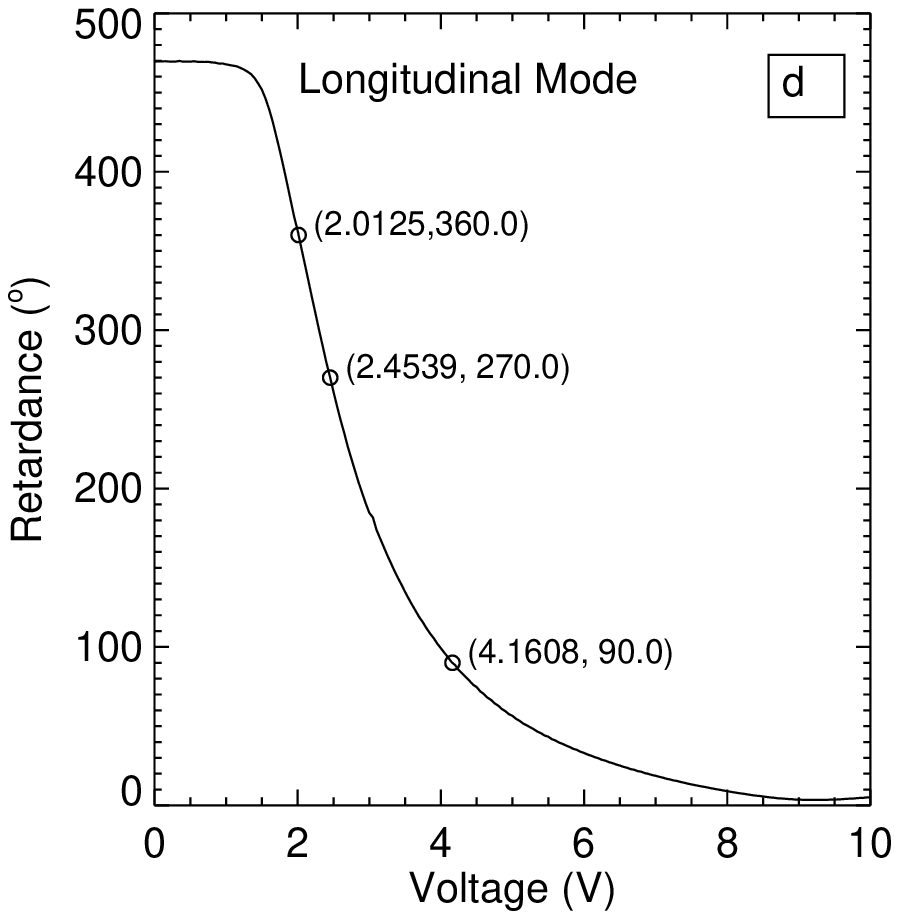}
             }
    \vspace{-0.35\textwidth}   % Shift close to the panel top 
    \centerline{\Large \bf     % Includes the labels (here needs the color package)
     \hspace{0.0 \textwidth} %\color{white}{}
     \hspace{0.415\textwidth}  %\color{white}{}
        \hfill}
    \vspace{0.31\textwidth}    % Shift back to the panel bottom 
             
\caption{Calibration curves of the LCVR1 (top panels) and LCVR2 (bottom panels) at 6173 \AA{} showing retardance as a function of voltage. Left (a and c) and right 
(b and d) panels show the voltages for the required retardance in vector and longitudinal modes, respectively. Vector mode will be used for the measurement 
of all the Stokes parameters and longitudinal mode will be used only for measuring Stokes parameters I and V.}
   %\label{fig6}
   \end{figure}	

   \begin{figure}    
   \centerline{\hspace*{0.015\textwidth}
               \includegraphics[width=0.515\textwidth,clip=]{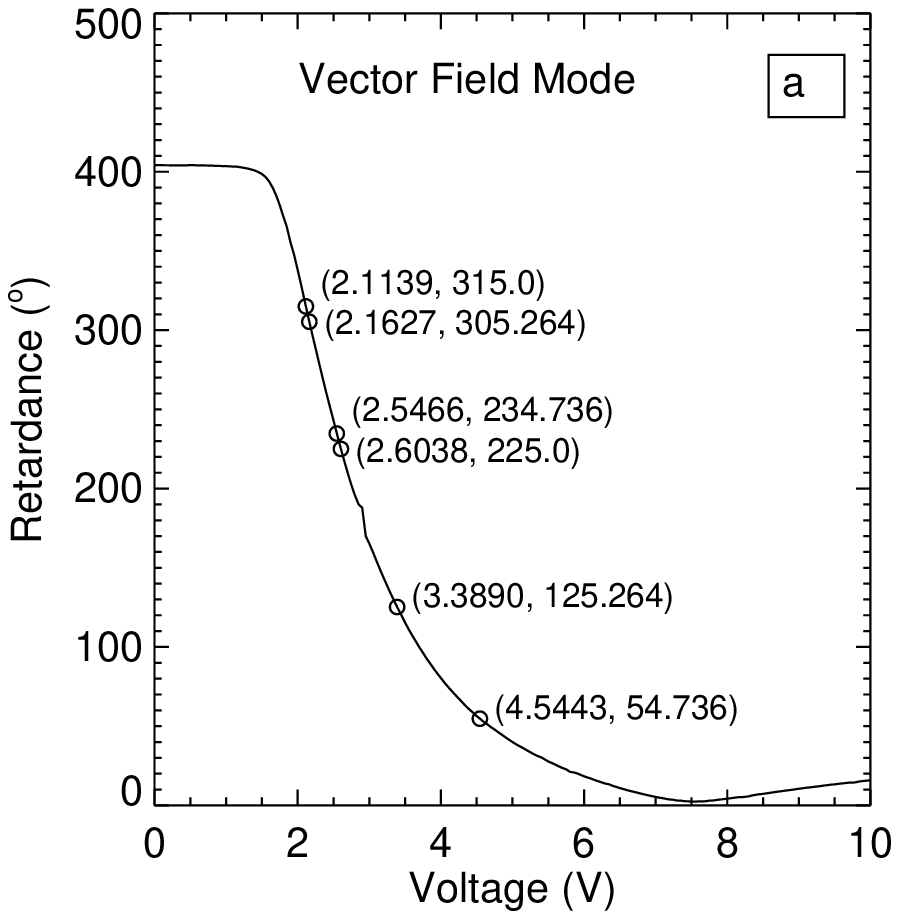}
               \hspace*{-0.03\textwidth}
               \includegraphics[width=0.515\textwidth,clip=]{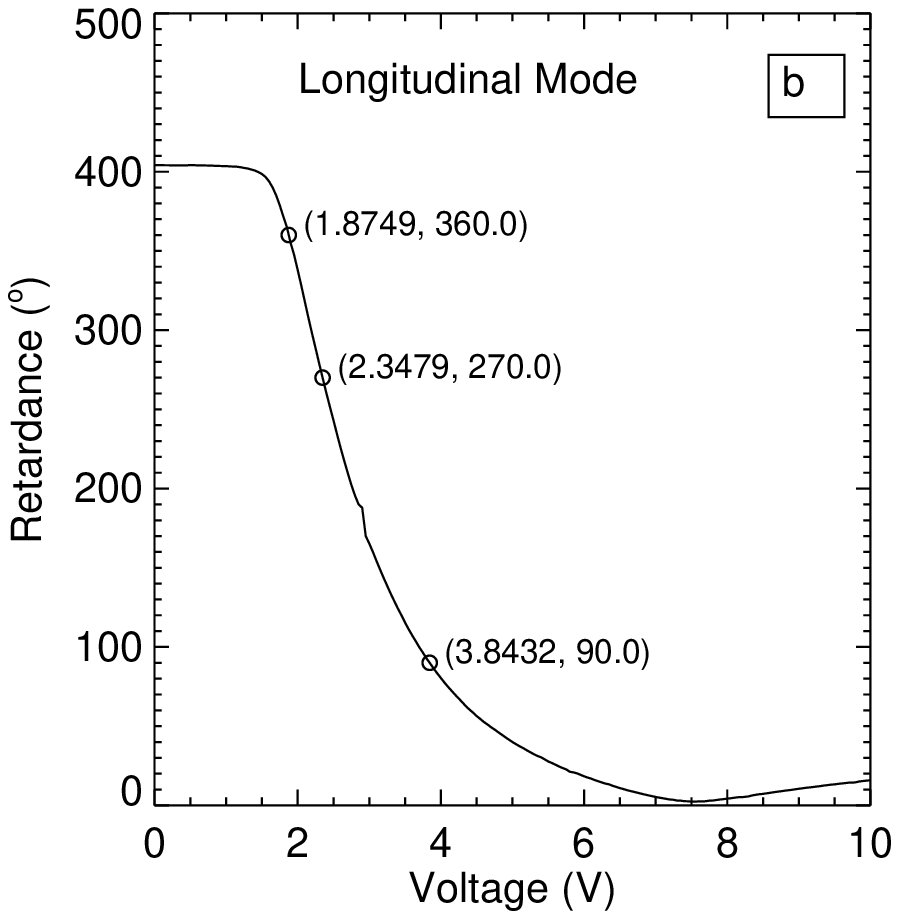}
              }
     \vspace{-0.35\textwidth}   % Shift close to the panel top 
     \centerline{\Large \bf     % Includes the labels (here needs the color 
                                %   package, see beginning of this file)
      \hspace{0.0 \textwidth}  %\color{white}{}
      \hspace{0.415\textwidth}  %\color{white}{}
         \hfill}
     \vspace{0.31\textwidth}    % Shift back to the panel bottom 
          
  \centerline{\hspace*{0.015\textwidth}
             \includegraphics[width=0.515\textwidth,clip=]{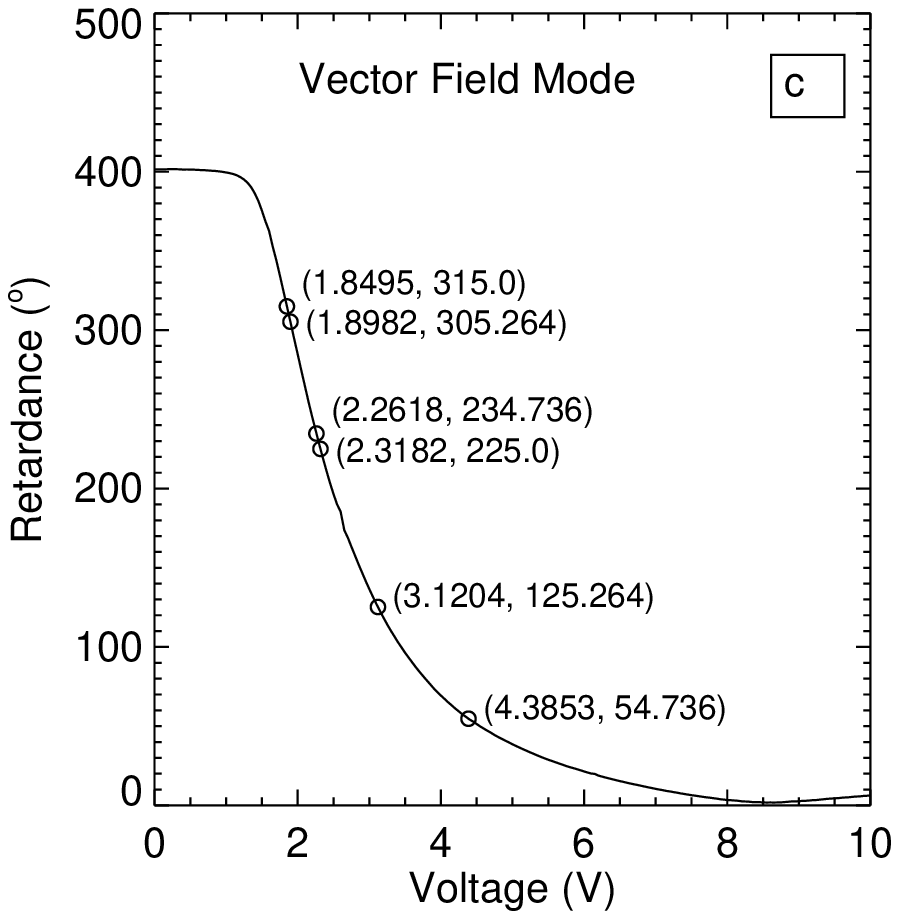}
             \hspace*{-0.03\textwidth}
              \includegraphics[width=0.515\textwidth,clip=]{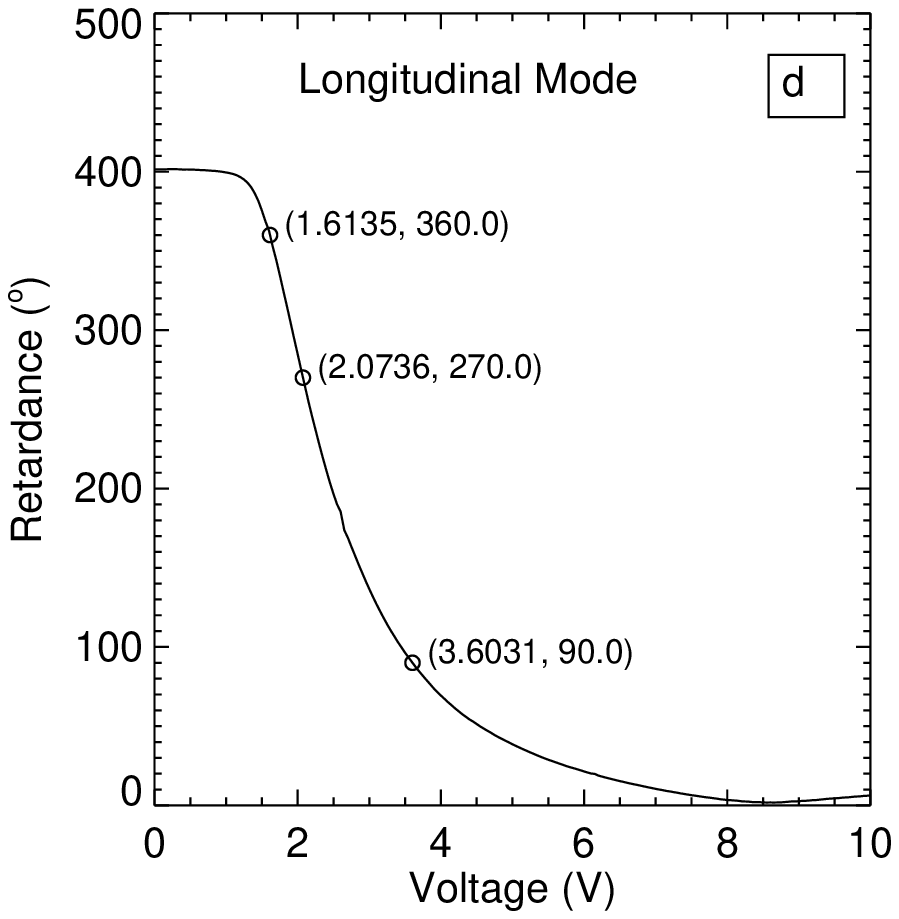}
             }
    \vspace{-0.35\textwidth}   % Shift close to the panel top 
    \centerline{\Large \bf     % Includes the labels (here needs the color package)
     \hspace{0.0 \textwidth} %\color{white}{}
     \hspace{0.415\textwidth}  %\color{white}{}
        \hfill}
    \vspace{0.31\textwidth}    % Shift back to the panel bottom 
             
\caption{Calibration curves of the LCVR1 (top panels) and LCVR2 (bottom panels) at 8542 \AA{} showing retardance as a function of voltage. Left (a and c) and right
(b and d) panels show the voltages for the required retardance in vector and longitudinal modes, respectively. Vector mode will be used for the measurement of 
all the Stokes parameters and longitudinal mode will be used only for measuring Stokes parameters I and V.}
        %\label{fig06}
   \end{figure}

\subsection{LCVRs: Voltage versus retardance} After knowing the crossed position of polarizers and the fast axis of LCVRs, we characterize the dependence 
of voltage on the retardance of the LCVRs using the following procedure. Keeping the polarizers P1 and P2 in crossed position and the fast axis of LCVR at 
$45^{\circ}$ with respect to P1, we applied voltages from $0$ to $10$ V in steps of $0.05$ V  to the LCVR. 
At each voltage five images were taken and computed the mean intensity to obtain $I_{out}^{90}$. Following the same procedure, 
$I_{out}^0$ as a function of voltage is obtained by rotating P2 such that P1 and P2 are in parallel position. 
In both the cases, the temperature of LCVR is kept constant at $28^{\circ}$C. 
\begin{table}
	%\centering
	\caption{Obtained voltages for the required retardance at 6173 \AA{} in the vector mode.}
%\label{tab4}
	\begin{tabular}{cccc}
		\hline
		$\delta_1$  & Voltage of LCVR1 & $\delta_2$  & Voltage of LCVR2\\
		($^{\circ}$) & (V) & ($^{\circ}$)  & (V)\\
		 \hline
		315.0 & 2.0977 & 305.264 & 2.2734\\ 
		315.0 & 2.0977 & 054.736 & 5.0480\\ 
	        225.0 & 2.5984 & 125.264 & 3.6152\\
	        225.0 & 2.5984 & 234.736 & 2.6513\\
		\hline
	\end{tabular}
        \end{table}
        
        \begin{table}
	%\centering
	\caption{Obtained voltages for the required retardance at 6173 \AA{} in the longitudinal mode.}
	%\label{tab5}
	\begin{tabular}{cccc}
		\hline
		$\delta_1$  & Voltage of LCVR1 & $\delta_2$  & Voltage of LCVR2\\
		($^{\circ}$) & (V) & ($^{\circ}$)  & (V)\\
		 \hline
	        360.0 & 1.8681 & 090.00 & 4.1608\\
	        360.0 & 1.8681 & 270.00 & 2.4539\\
		\hline
	\end{tabular}
        \end{table}
With the measured $I_{out}^0$ and $I_{out}^{90}$ at each voltage, the retardance is calculated using Equation (22).
The retardance of LCVRs (LCVR1 and LCVR2) with voltage for 6173 \AA{} and 8542 \AA{} are plotted and shown in Figure 6 and 7, 
respectively. The characteristics plots between retardance and voltage are used to estimate the voltages required in both the modulation schemes 
(vector and longitudinal). 
Figures 6 (a) and 6 (c) represent the corresponding voltages required for the retardance in vector mode modulation 
scheme (shown by circle on the characteristic curve) and Figures 6 (b) and 6 (d) represent the corresponding voltages required 
for the retardance in longitudinal mode modulation scheme, for LCVR1 and LCVR2 at wavelength 6173 \AA{}, respectively. 
\newline
Similarly, panels (a) and (c) of Figure 7 represent corresponding voltages required for the retardance in vector mode modulation 
scheme and panels (b) and (d) of Figure 7 represent corresponding voltages required for the retardance in longitudinal mode modulation 
scheme for LCVR1 and LCVR2 at wavelength 8542 \AA{}, respectively. 
\newline
The retardance and their corresponding voltages calibrated from Figures 6 and 7 are listed in Table $1-4$ for both the LCVRs 
according to their respective wavelengths. These values are used for the measurement of Stokes parameters with the polarimeter for MAST.
        \begin{table}
	%\centering
	\caption{Obtained voltages for the required retardance at 8542 \AA{} in the vector mode.}
        %\label{tab6}
	\begin{tabular}{cccc}
		\hline
		$\delta_1$  & Voltage of LCVR1 & $\delta_2$  & Voltage of LCVR2\\
		($^{\circ}$) & $(V)$ & ($^{\circ}$)  & $(V)$\\
		 \hline
		315.0 & 2.1139 & 305.264 & 1.8982 \\ 
		315.0 & 2.1139 & 054.736 & 4.3853\\ 
	        225.0 & 2.6038 & 125.264 & 3.1203\\
	        225.0 & 2.6038 & 234.736 & 2.2618\\
	        \hline
	        \end{tabular}
                \end{table}
                
	 \begin{table}
	\caption{Obtained voltages for the required retardance at 8542 \AA{} in the longitudinal mode.}
       %\label{tab7}
	\begin{tabular}{cccc}
		\hline
		$\delta_1$  & Voltage of LCVR1 & $\delta_2$  & Voltage of LCVR2\\
		($^{\circ}$) & $(V)$ & ($^{\circ}$)  & $(V)$\\
		 \hline
	        360.0 & 1.8749 & 090.00 & 3.6031\\
	        360.0 & 1.8749 & 270.00 & 2.0736\\
		\hline
	\end{tabular}
        \end{table}

        \begin{figure}
    \centerline{\includegraphics[width=0.50\textwidth,clip=]{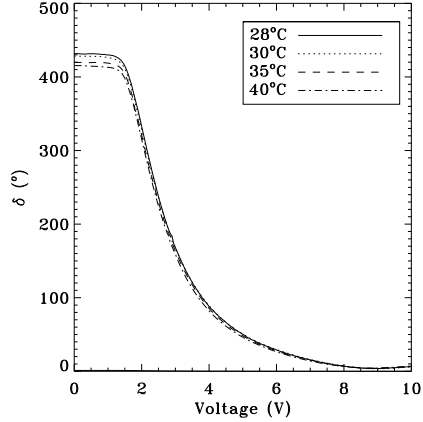}
    }
    \caption{Variation of the retardance with the applied voltage of LCVR for four different temperatures. It is evident that as temperature 
    increases the retardance of the LCVR decreases.}
    %\label{fig7}
\end{figure}

\subsection{LCVRs: Temperature versus retardance} As evident from Equation (14) that the temperature also influences retardance of LCVRs, we have 
characterized voltage dependence of retardance of a LCVR at four different temperatures, \textit{i.e.}, $28^{\circ}$C, $30 ^{\circ}$C, $35^{\circ}$C, and 
$40 ^{\circ}$C. Figure 8 shows a change in retardance of LCVR with the voltage  at different temperatures. As evident from Figure 8, retardance of the 
LCVR decreases as their temperature increases. The effect of temperature on the retardance of the LCVR can be clearly seen in Figure 9. 
It shows that the LCVR is more sensitive to the temperature when it is operated at low voltages $(0-4 V)$ \citep{Li2004, Capobianco2008}. 
Maximum change in the retardance with the temperature of $\sim-1.5^{\circ}$/$^{\circ}$C is observed at $2.0$ V.
At present the LCVRs are kept in a temperature enclosures provided by the vendor. Temperature stability of the enclosure is $\pm 1^{\circ}$C.
\newline
For precise polarimetric measurements, it is important to know the change in response matrix due to the fluctuations in the temperature enclosure. 
As mentioned above, $\pm1.0^{\circ}$C variation in temperature causes a maximum change in retardance of $-$1.5$^{\circ}$. 
Incorporating the change in retardance, the new response matrix can be written as,
\begin{equation}%\label{eqn23}
\textbf{X \textprime}
=
\begin{pmatrix}
0.99983 & 0.00017 & 0.01309 & -0.01309\\ 
0.00017 & 0.99983 & -0.01309 & 0.01309\\
-0.01309 & 0.01309 & 0.99983 & 0.00017\\
0.01309 & -0.01309 & 0.00017 & 0.99983
\end{pmatrix}.
\end{equation}
Thus, the error in the measurement of response matrix is, 
$$
\Delta \textbf{X}=\textbf{X}-\textbf{X\textprime}=
\begin{pmatrix}
0.00017 & -0.00017 & -0.01309 & 0.01309\\ 
-0.00017 & 0.00017 & 0.01309 & -0.01309\\
0.01309 & -0.01309 & 0.00017 & -0.00017\\
-0.01309 & 0.01309 & -0.00017 & 0.00017
\end{pmatrix},
$$
where X is unity matrix. The matrix element (3, 2), (4, 2), (4, 3) of $X\textprime$ shows the cross-talk among the Stokes parameters Q, U, and V. 
In this case, the cross-talk from Q to U and Q to V is same and equal to $1.3\times10^{-2}$ and U to V cross-talk is $1.7\times10^{-4}$. As the expected 
polarimetric accuracy is poorer, we calculated the change in polarimetric accuracy for different value of temperature accuracy. Table 8 summarises the calculations.
 For a temperature variation of $\pm 0.25^{\circ}$C (which can be obtained by optimizing the temperature control system), causes an error in 
retardance of $\pm 0.38^{\circ}$ (as shown in Table 8), the cross-talk will be in the order of $10^{-3}$ which would be acceptable for our scientific studies. 
Therefore a temperature control system which can maintain the temperature of LCVRs within $\pm 0.25^{\circ}$C will be constructed and used for the 
measurements of the Stokes parameters.
\begin{table}
	\caption{Polarimetric cross-talk due to change in the temperature stability ($\delta T$) of the temperature enclosure.}
       %\label{tab7}
	\begin{tabular}{cccc}
		\hline
                $\delta T$  & Q $\rightarrow$ U & Q$\rightarrow$ V & $U \rightarrow V$\\
		($^{\circ}$C) &cross-talk &cross-talk &cross-talk \\
		 \hline
	        1.00 & $1.3 \times 10^{-2}$ &$1.3 \times 10^{-2}$  & $1.7 \times 10^{-4}$\\
	        0.50 & $6.5 \times 10^{-3}$ & $6.5 \times 10^{-3}$ & $4.25 \times 10^{-5}$\\
	        0.25 & $3.2 \times 10^{-3}$ & $3,2 \times 10^{-3}$ & $1.04 \times 10^{-5}$\\
		\hline
	\end{tabular}
        \end{table}

\begin{figure}   
                                % includes the two top panels 
   \centerline{\hspace*{0.015\textwidth}
               \includegraphics[width=0.515\textwidth,clip=]{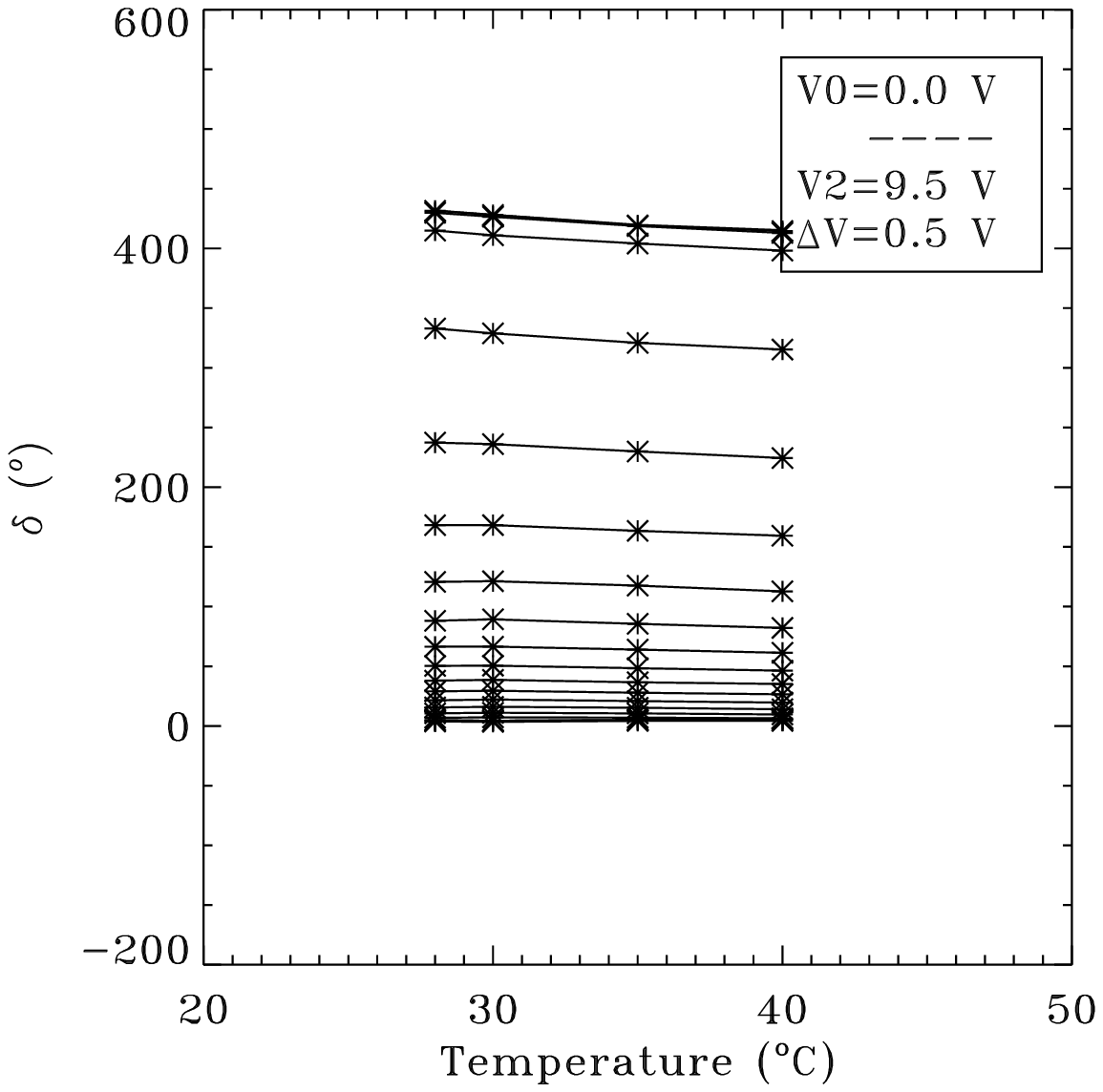}
               \hspace*{-0.03\textwidth}
               \includegraphics[width=0.515\textwidth,clip=]{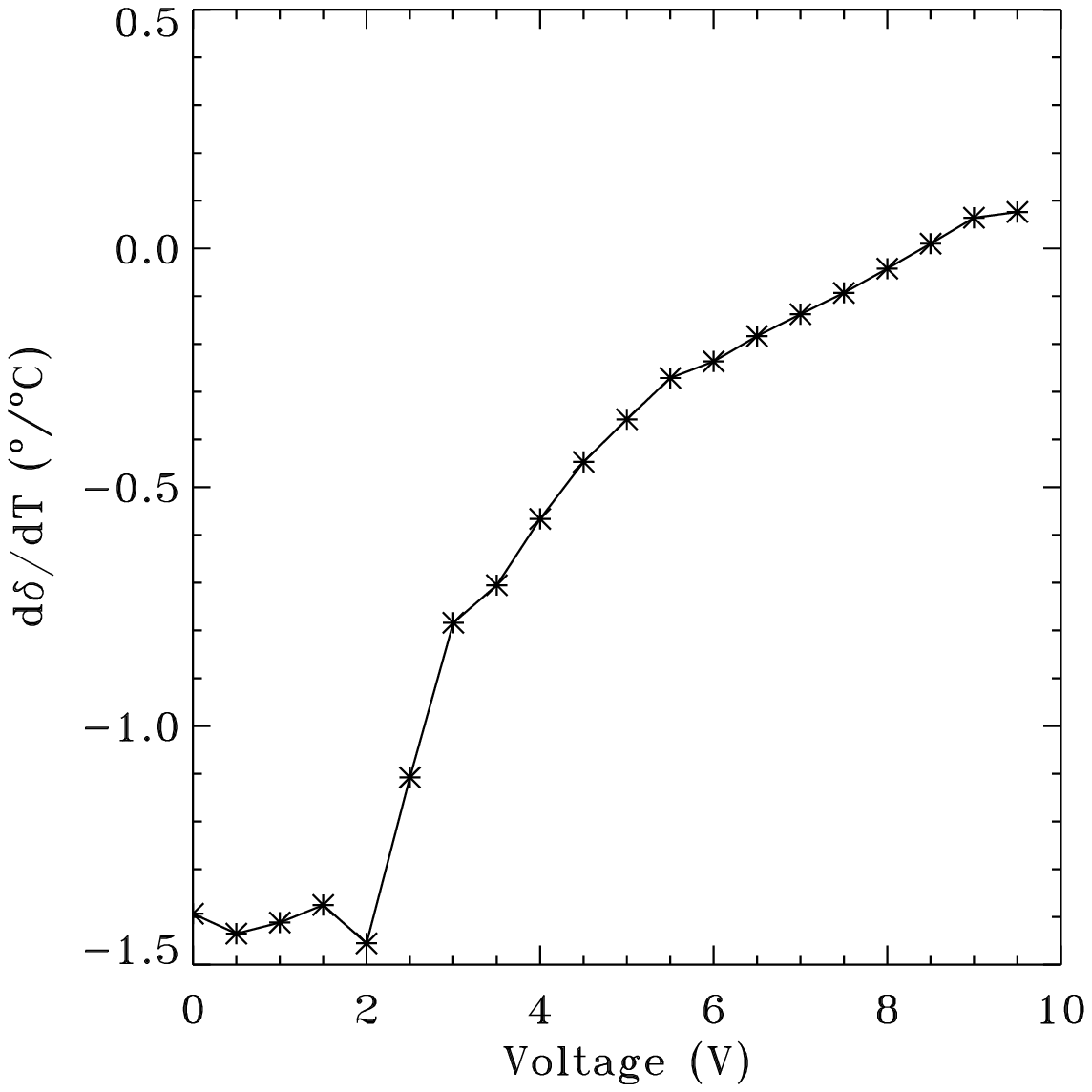}
              }
     \vspace{-0.35\textwidth}   % Shift close to the panel top 
     \centerline{\Large \bf     % Includes the labels (here needs the color 
                                %   package, see beginning of this file)
      \hspace{0.0 \textwidth}  \color{white}{}
      \hspace{0.415\textwidth}  \color{white}{}
         \hfill}
     \vspace{0.31\textwidth}    % Shift back to the panel bottom 
    
      \caption{Left: Change in retardance with the change in temperature for different voltages. Right: retardance derivative with respect to the temperature 
      as a function of voltage which shows that LCVR is insensitive to temperature for higher voltages.}
   %\label{fig007}
   \end{figure}	

\subsection{LCVRs: Change in the orientation of the fast axis with the voltage} It is presumed that the angular position of the LCVR 
fast axis is independent of the voltage, only the retardation changes according to the voltage. But in practice, it is observed that the position of the fast 
axis also changes with the voltage \citep{Terrier2010}. In order to see the effect of voltage on the orientation of the fast axis of LCVR, 
we performed the following experiment. Two polarizers P1 and P2 are placed in a collimated beam keeping P1 at a reference position and intensity is 
measured by rotating P2 from $0^{\circ}-180^{\circ}$. After knowing the crossed position of the polarizers P1 and P2, LCVR is placed between P1 and P2. 
The orientation of LCVR is adjusted such that the fast axis of LCVR becomes parallel to polarizer P1 (Figure 10).
In this configuration, rotating the P2 gives exactly the same kind of intensity variation as in the case of crossed linear polarizers. Without changing the 
orientation of the LCVR, we apply different voltages (between 0-10 V) to LCVR and measured the intensity variation by
rotating P2. We observed that (Figure 10, left) the intensity variation is sinusoidal again, and the sinusoid pattern for different voltages is 
shifted with respect to the reference sinusoid (no LCVR). The difference between the reference position (no LCVR or without voltage) and the
actual position for different voltages is measured and plotted against the applied voltage for LCVR1 (see Figure 10, right).

\begin{figure}   
                                % includes the two top panels 
   \centerline{\hspace*{0.015\textwidth}
               \includegraphics[width=0.515\textwidth,clip=]{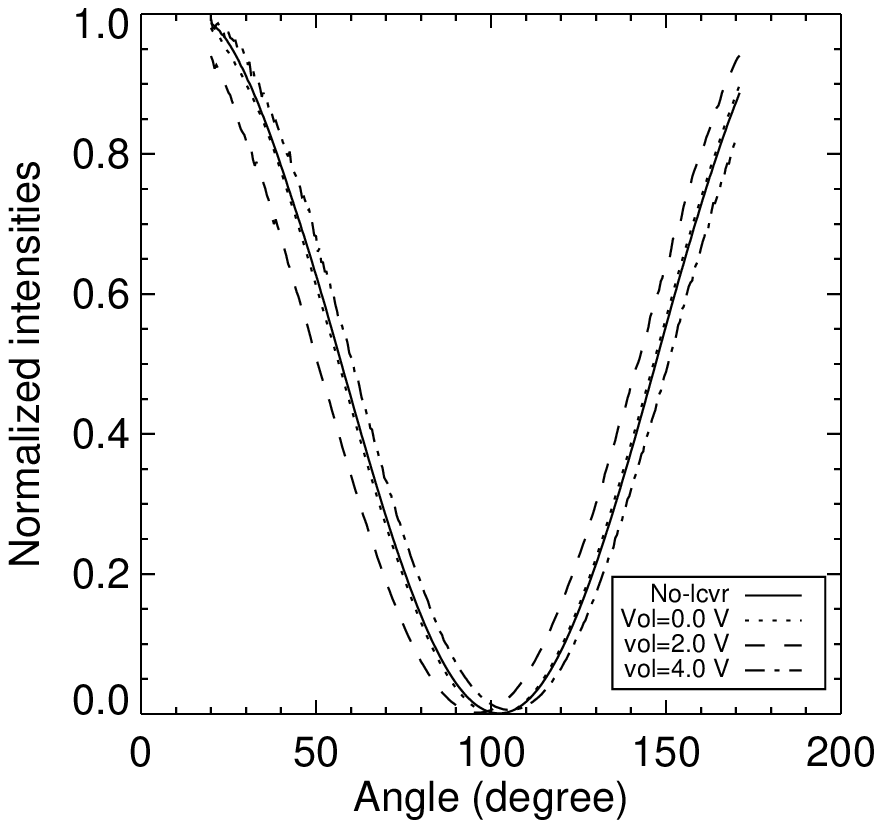}
               \hspace*{-0.03\textwidth}
               \includegraphics[width=0.515\textwidth,clip=]{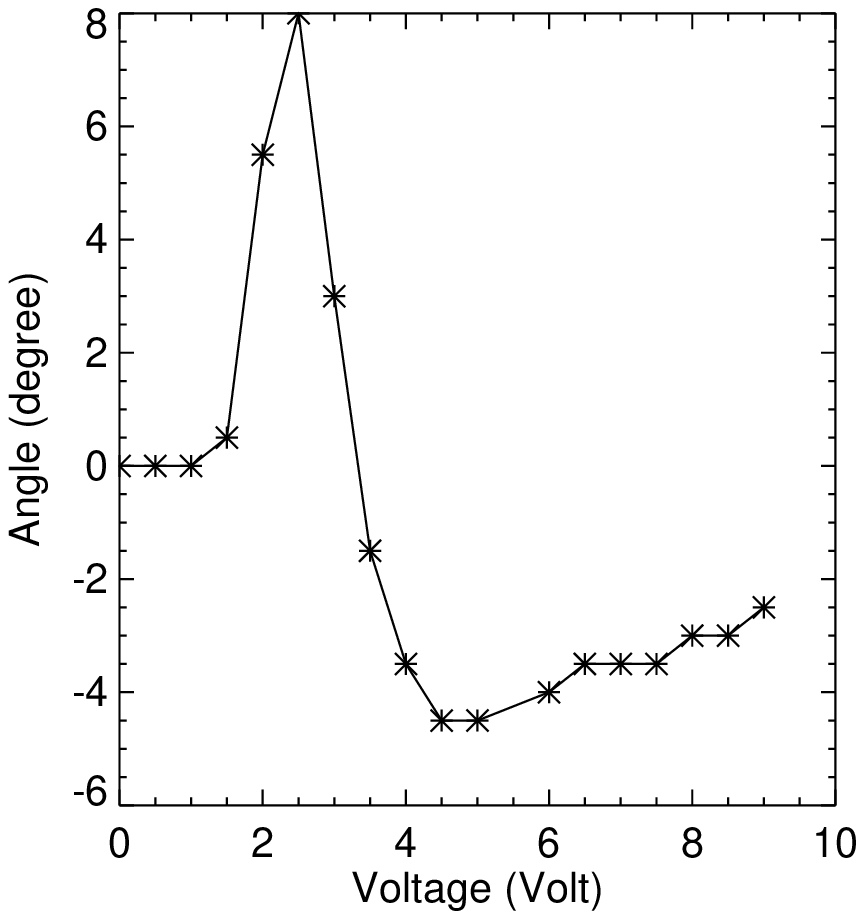}
              }
     \vspace{-0.35\textwidth}   % Shift close to the panel top 
     \centerline{\Large \bf     % Includes the labels (here needs the color 
                                %   package, see beginning of this file)
      \hspace{0.0 \textwidth}  \color{white}{}
      \hspace{0.415\textwidth}  \color{white}{}
         \hfill}
     \vspace{0.31\textwidth}    % Shift back to the panel bottom 
    
      \caption{Left panel shows the variation of intensity with the change in angular position of polarizer P2 with reference to polarizer P1 for different 
      voltages applied to the LCVR. The right panel shows the shift in the position of the fast axis with the voltage applied to the LCVR in which reference is 
      taken as no LCVR position.}
   %\label{fig101}
   \end{figure}

 As shown in Figure 10 (right), the maximum shift obtained in the orientation of the fast axis of LCVR is $8^{\circ}$ at $2.5$ V. 
 Thus, it is important to know the change in response matrix due to the shift in the orientation of fast axis.  The new response matrix due to the change in the 
 orientation of the fast axis ($8^{\circ}$) is,

 \begin{equation}%\label{eqn24}
\textbf{X\textprime}
=
\begin{pmatrix}
0.99454 & 0.03327 & 0.05467 & -0.08248\\ 
0.03327 & 0.99454 & 0.05467 &-0.08248\\
0.30782 & -0.20404 & 0.92874 & -0.03252\\
0.30782 & -0.20404 & -0.03252 & 0.92874
\end{pmatrix}.
\end{equation}
The error in the measurement of modulation matrix is, 
$$
\Delta \textbf{X}=\textbf{X}-\textbf{X\textprime}=
\begin{pmatrix}
0.00546 & -0.03327 & -0.05467 & 0.08248\\ 
-0.03327 & 0.00546 & -0.05467 & 0.08248\\
-0.30782 & 0.20404 & 0.07126 & 0.03252\\
-0.30782 & 0.20404 & 0.03252 & 0.07126
\end{pmatrix},
$$
where X is unity matrix. The matrix element (3, 2), (4, 2), (4, 3) of $X\textprime$ shows the cross talk among the Stokes parameters Q, U, and V. 
In this case the cross-talk from Q to U and Q to V is same and equal to $2.0\times10^{-1}$ and U to V 
cross-talk is $3.2\times10^{-2}$. 

As evident from the above analysis the cross-talk in the Stokes measurement resulting from the drift in the LCVR fast axis while applying voltages is 
considerably large and need to be taken care of. The theoretical modulation matrix O (Equation (9)) is modified by including the drift in the fast axis 
for corresponding voltages. This modified modulation matrix is used for the demodulation of the observed Stokes profiles \citep{Terrier2010}.

\section{Experimental determination of response matrix of the polarimeter}
 As discussed in Section 2.1, the relation between the incoming Stokes vector and measured Stokes vector can be written as 
 \begin{equation}
S_{meas}=\textbf{X}.S_{in},
\end{equation}
where \textbf{X} is the $4\times4$ element response matrix. The response matrix of the polarimeter can be determined experimentally from the measured Stokes 
parameters of the known input polarizations generated by calibration unit consisting of a zero order quarter wave plate (QWP) and a linear polarizer. 
We have computed the response matrix of the polarimeter in the laboratory using experimental setup shown in Figure 11 for both the wavelengths (6173 \AA{} and 
8542 \AA{}).\newline

 \begin{figure}
	\centerline{\includegraphics[width=0.98\textwidth,clip=]{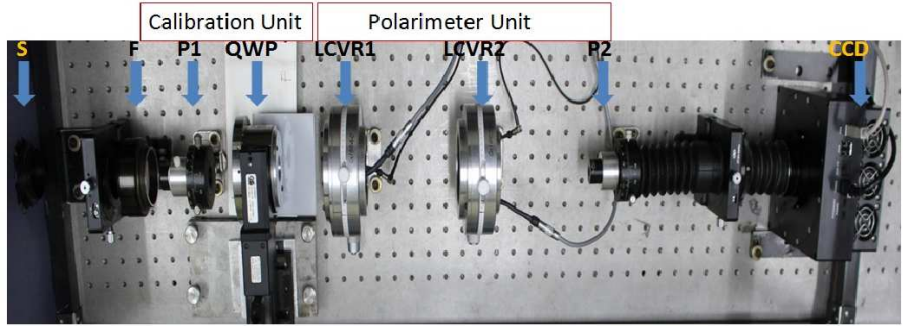}
	}
	\caption{Experimental setup for the calibration of response matrix for MAST polarimeter. In this setup, a light beam is collimated using lens L1 and then 
	collimating light passes through an interference filter (F), calibration unit consists of a linear polarizer (P1) and a zero order quarter wave plate (QWP), 
	and polarimeter consists of two LCVRs (LCVR1 and LCVR2) and a linear polarizer (P2). 
	Finally, image is formed by imaging lens on CCD which is placed in the focal plane of the imaging lens.}
	%\label{fig8}
\end{figure}
For the response matrix calibration, the calibration unit (CU) is placed in the beam before the polarimeter module as shown in the
Figure 8. The orientation of the axes has been determined with an accuracy of $±1^{\circ}$ for the linear polarizer, and $±0.5^{\circ}$ 
for the QWP of CU which has been fixed in a computer controlled rotating mount. To determine \textbf{X} of the polarimeter, 
80 known polarization states are created by rotating QWP from $0^{\circ}-160^{\circ}$ with a step size of $2^{\circ}$ and measured by the 
polarimeter using four intensity measurement modulation scheme (Table 1) and six measurement modulation scheme (Table 2) as discussed in Section 2.1. 
% \newline
We have computed the response matrix of the polarimeter using both the schemes. 
Here, we discuss the response matrix calibration using four measurement 
modulation scheme in detail. 
\begin{figure}    
   \centerline{\hspace*{0.015\textwidth}
               \includegraphics[width=0.515\textwidth,clip=]{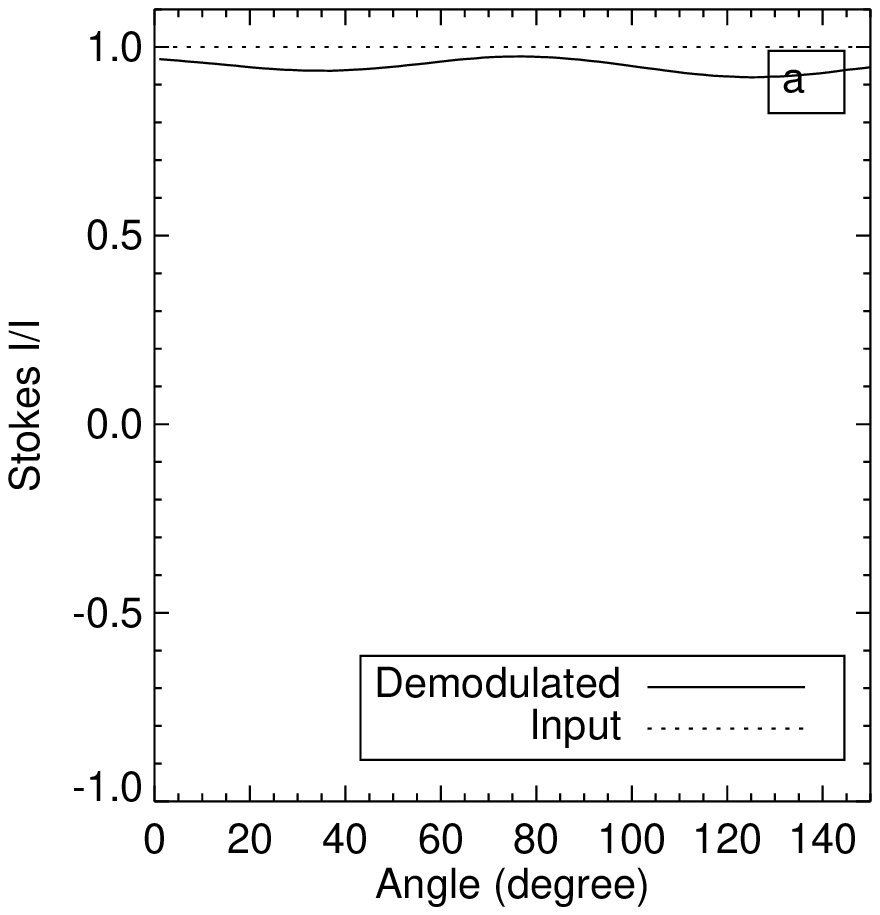}
               \hspace*{-0.03\textwidth}
               \includegraphics[width=0.515\textwidth,clip=]{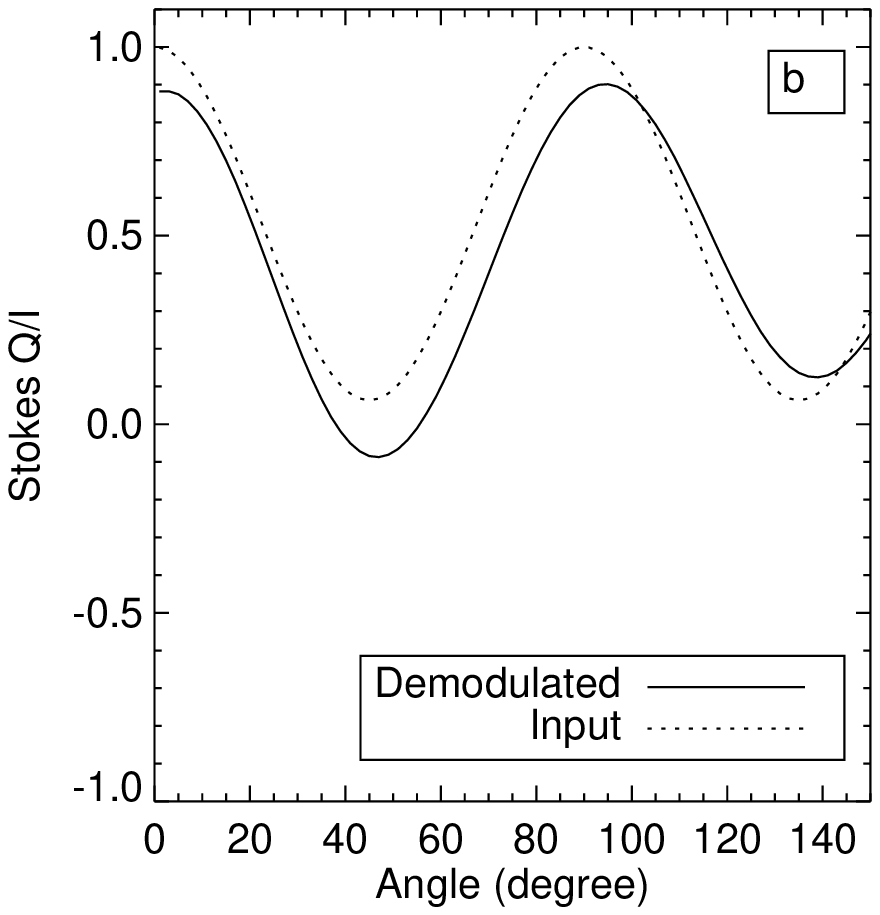}
              }
     \vspace{-0.35\textwidth}   % Shift close to the panel top 
     \centerline{\Large \bf     % Includes the labels (here needs the color 
                                %   package, see beginning of this file)
      \hspace{0.0 \textwidth}  %\color{white}{}
      \hspace{0.415\textwidth}  %\color{white}{}
         \hfill}
     \vspace{0.31\textwidth}    % Shift back to the panel bottom 
          
  \centerline{\hspace*{0.015\textwidth}
             \includegraphics[width=0.515\textwidth,clip=]{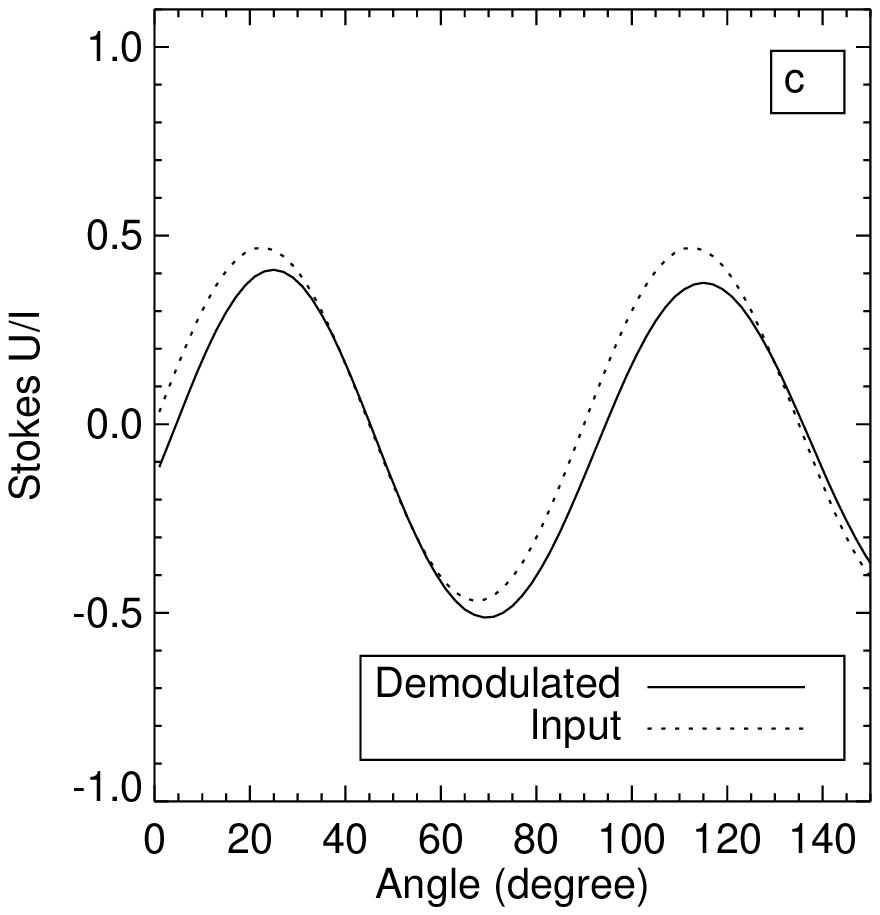}
             \hspace*{-0.03\textwidth}
              \includegraphics[width=0.515\textwidth,clip=]{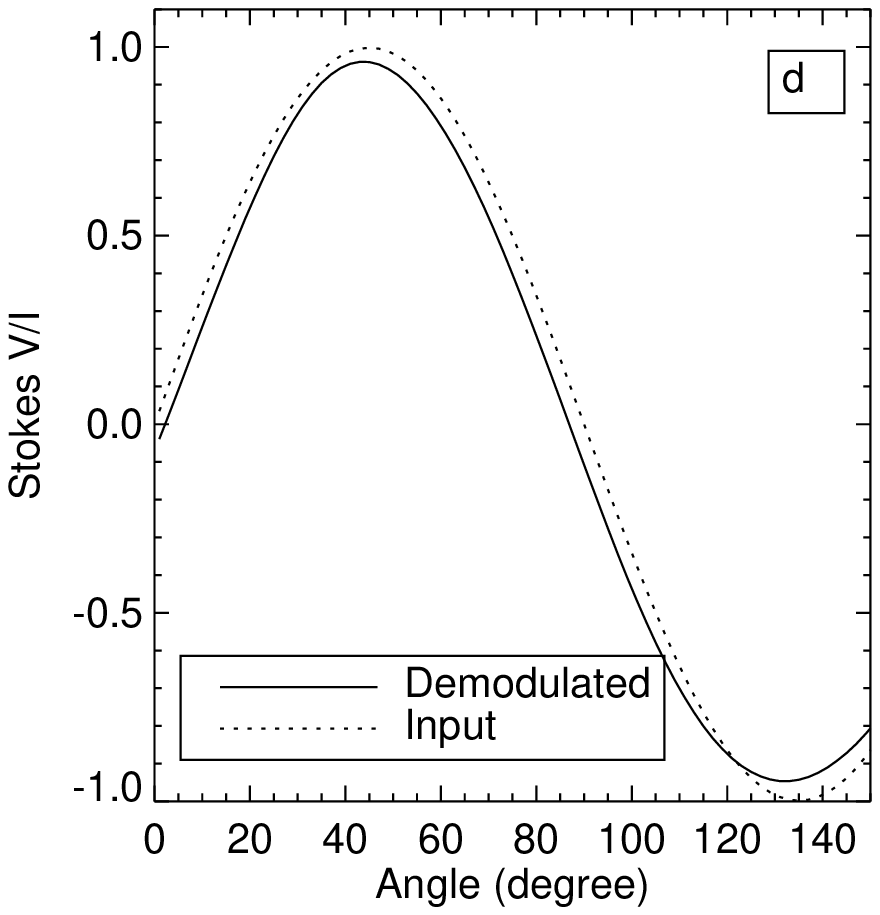}
             }
    \vspace{-0.35\textwidth}   % Shift close to the panel top 
    \centerline{\Large \bf     % Includes the labels (here needs the color package)
     \hspace{0.0 \textwidth} %\color{white}{}
     \hspace{0.415\textwidth}  %\color{white}{}
        \hfill}
    \vspace{0.31\textwidth}    % Shift back to the panel bottom 
             
\caption{Plots of input Stokes parameters and demodulated Stokes parameters calculated at each position angles 
of QWP of CU for 6173 \AA{} wavelength.}
        %\label{fig9}
   \end{figure}
For the configuration shown in Figure 11, the input Stokes vector can be written as,
  \begin{figure}    
   \centerline{\hspace*{0.015\textwidth}
               \includegraphics[width=0.515\textwidth,clip=]{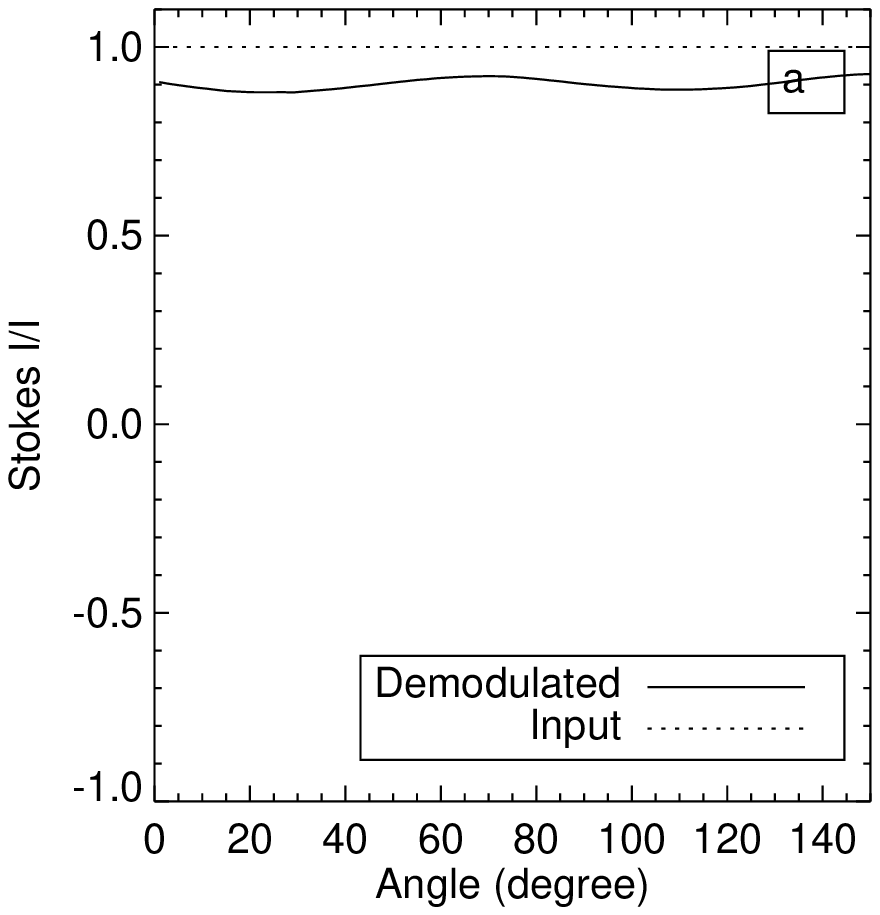}
               \hspace*{-0.03\textwidth}
               \includegraphics[width=0.515\textwidth,clip=]{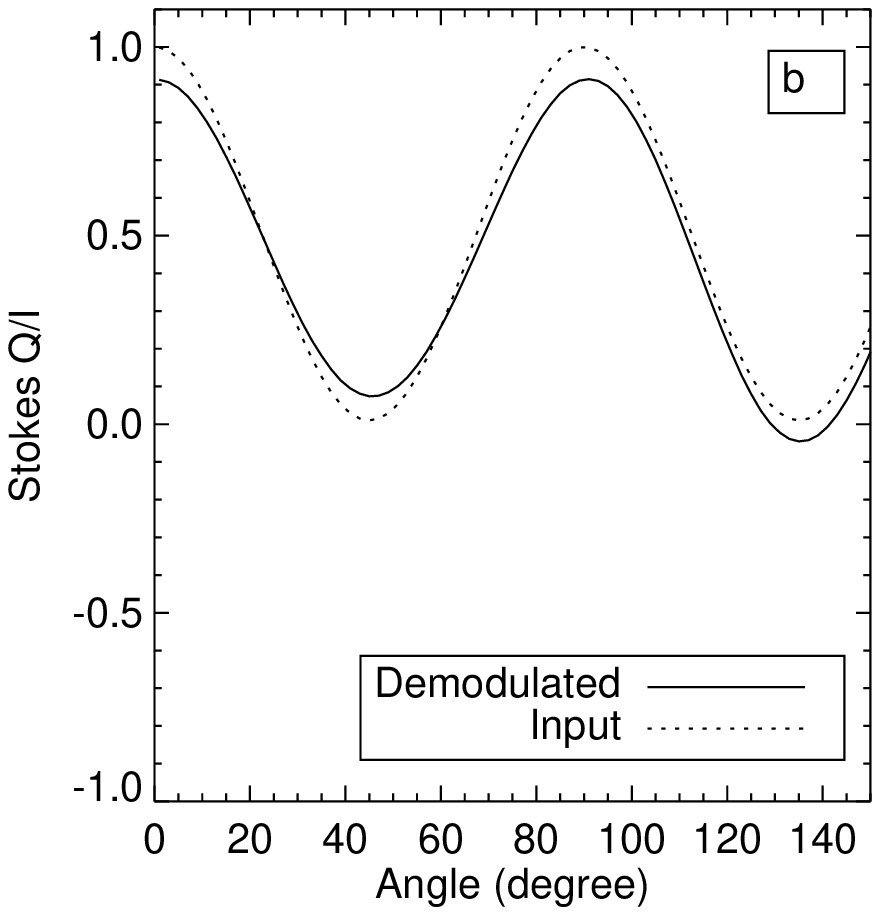}
              }
     \vspace{-0.35\textwidth}   % Shift close to the panel top 
     \centerline{\Large \bf     % Includes the labels (here needs the color 
                                %   package, see beginning of this file)
      \hspace{0.0 \textwidth}  %\color{white}{}
      \hspace{0.415\textwidth}  %\color{white}{}
         \hfill}
     \vspace{0.31\textwidth}    % Shift back to the panel bottom 
          
  \centerline{\hspace*{0.015\textwidth}
             \includegraphics[width=0.515\textwidth,clip=]{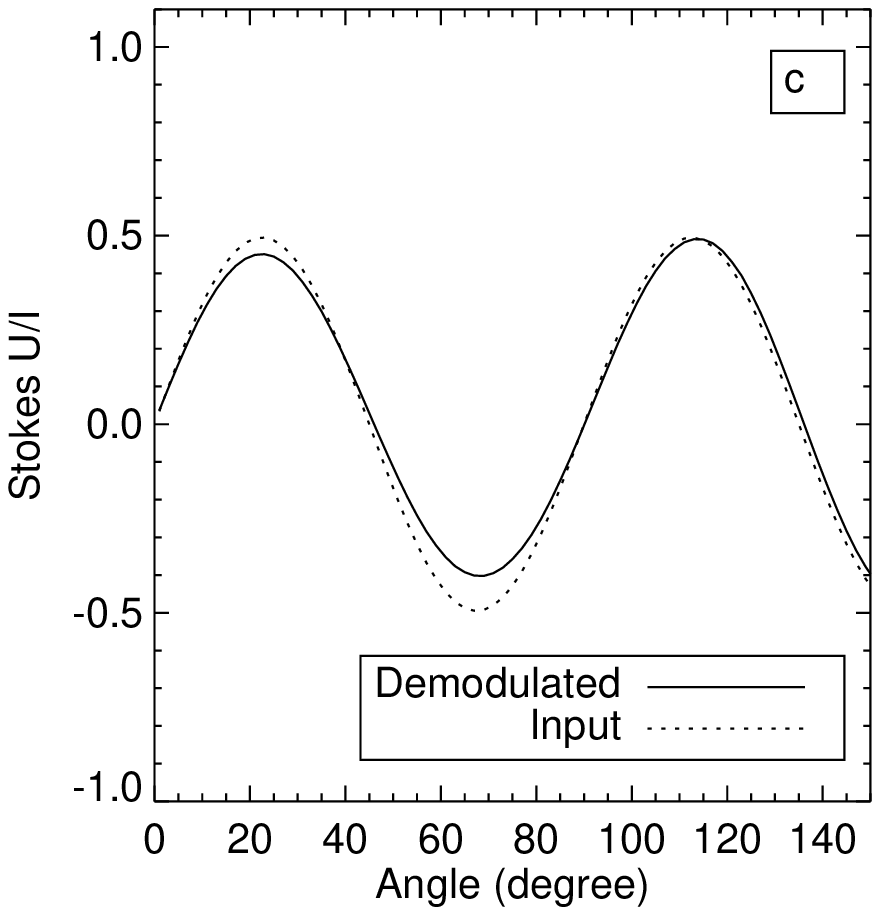}
             \hspace*{-0.03\textwidth}
              \includegraphics[width=0.515\textwidth,clip=]{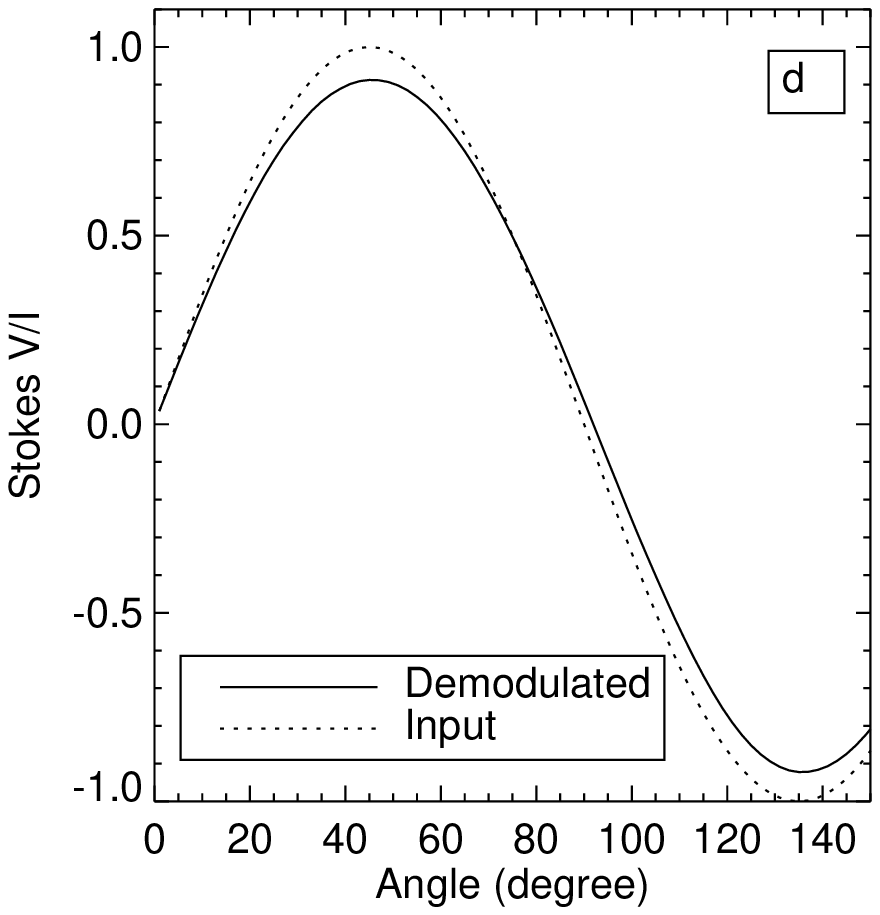}
             }
    \vspace{-0.35\textwidth}   % Shift close to the panel top 
    \centerline{\Large \bf     % Includes the labels (here needs the color package)
     \hspace{0.0 \textwidth} %\color{white}{}
     \hspace{0.415\textwidth}  %\color{white}{}
        \hfill}
    \vspace{0.31\textwidth}    % Shift back to the panel bottom 
             
\caption{Plots of input Stokes parameters and demodulated Stokes parameters calculated at each position angles of QWP of CU for 8542 \AA{} wavelength.}
        %\label{fig10}
   \end{figure}
$$
S_{in}=M_{QWP}(\theta_r,\delta).M_{P1}(\theta_p).[1 0 0 0]^T, 
$$
where, $\theta_r$ and $\theta_p$ are the orientation of the retarder and polarizer of the CU relative to the reference axis 
and $\delta$ is the retardance of QWP. In our case, we fixed $\theta_p$ at $0^{\circ}$, then the input Stokes parameters for 
different retarder orientation is given by, 
\begin{equation}
 \left.\begin{aligned}
        I=1, \\
        Q=cos^2(2\theta)+sin^2(2\theta).cos(\delta),\\
        U=sin(2\theta).cos(2\theta).(1-cos(\delta)),\\
        V=-sin(2\theta).sin(\delta). 
        \end{aligned}
 \right\}
\end{equation}
For \textit{n} orientations of CU retarder, polarimeter response matrix is calibrated from the measurements after 
rearranging the Stokes vector into n$\times$4 matrices by a solution of the linear problem,
$$
S_{meas}=X.S_{in},
$$
Multiplying by $S_{in}^T$ from the right in above Equation
$$
S_{meas}.S_{in}^T=X.S_{in}.S_{in}^T=X.D,
$$
where
$$
D=S_{in}.S_{in}^T.
$$
Therefore, the final expression for response matrix is
\begin{equation}
X=S_{meas}.S_{in}^T.D^{-1} .
\end{equation}
Hence, response matrix for 6173 \AA{} is determined from the above Equation using input and measured Stokes vector is given by
  \begin{equation}
X_4^{6173} = 
\begin{pmatrix}
1.0000& -0.0507 & -0.0054 &-0.0595\\
-0.0090 & 0.8946 & -0.1618 &-0.0369\\
-0.0447 & -0.0639 & 0.8314 &0.0734\\
-0.0474 & -0.1494 & 0.0647 &0.9665\\
\end{pmatrix}
\end{equation}
Thus, the real incoming Stokes vector $\vec{S_{in}}$ can be calculated from the observed Stokes vector and measured 
response matrix as follows,
$$
\vec{S_{in}}=X^{-1}.S_{meas} .
$$
The input and the demodulated Stokes parameters at each CU retarder orientation are shown in Figure 12.
\newline
Similarly, we have determined the response matrix of the polarimeter at 8542 \AA{} wavelength and given by, 
 \begin{equation}%\label{eqn25}
X_4^{8542} = 
\begin{pmatrix}
1.0000& 0.0006 & 0.0349 &-0.0965\\
-0.0009 & 0.8931 & -0.1053 & 0.0041\\
-0.0135 & -0.0616 & 0.9073 & -0.1809\\
0.0873 & 0.0452 & 0.1948 &0.9115\\
\end{pmatrix},
\end{equation}
and plots for the input and demodulated input Stokes parameters are shown in Figure 13. 
	\begin{figure}
\centerline{\includegraphics[width=0.98\textwidth,clip=]{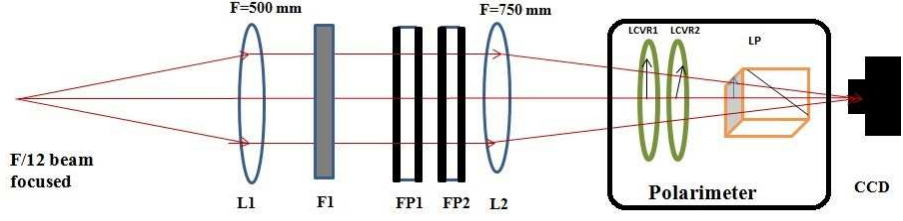}
           }
	\caption{Schematic diagram of the imaging polarimeter for MAST. In this setup a F/12 beam from the 
	telescope is collimated using the lens L1. The collimating beam passes through narrow-band imager consists of two Fabry-Perot 
	etalons (FP1 and FP2) and prefilter F1. Polarimeter consist of two LCVR and a Glan-Thompson polarizer is kept in between 
	the CCD and imaging lens (L2) in the converging beam. The F-number of the converging beam is 18.} 
	%\label{fig_15}
\end{figure}

\subsection{Response matrix for six measurement modulation scheme} Similarly, we have also computed X when Stokes parameters were obtained by six measurement 
modulation scheme and the other procedures were same as discussed above.
The response matrix of the polarimeter for 6173 \AA{} is,
  \begin{equation}%\label{eqn26}
X_6^{6173}= 
\begin{pmatrix}
1.0000 & -0.0084 & 0.0153 & -0.0016\\
0.0009 & 0.9493 & 0.0265 & 0.1462\\
-0.0171 & 0.0254 & 0.9573 & 0.9491\\
0.0043& -0.1829 &-0.0808 & 0.9491\\
\end{pmatrix}.
\end{equation}
Similarly, the response matrix of the polarimeter for 8542 \AA{} is,
\begin{equation}%\label{eqn27}
X_6^{8542}= 
\begin{pmatrix}
1.0000 & 0.0062 & -0.0177 & 0.0044\\
-0.0056 & 0.9251 & 0.0665 & 0.1308\\
-0.0652 & 0.0396 & 0.9454 & 0.0791\\
0.0344& -0.1379 &-0.1509 & 0.9176\\
\end{pmatrix}.
\end{equation}

\section{Preliminary observations of Stokes profiles}
Preliminary observations were obtained using the imaging spectropolarimeter for MAST, consisting of a narrow-band imager and a polarimeter.  
The narrow-band imager is designed around two z-cut $LiNbO_3$ Fabry-Perot etalons placed in tandem to provide a better spectral resolution. 
The wavelength characterization of the components of the narrow band imager is described in a previous paper \citep{RajaBayanna2014}. The integration and 
test results of the imager obtained with MAST will be presented in a separate paper (Mathew et al., (in preparation)). The full-width at half maximum 
(FWHM) of the FP combination is $\sim$95 m\AA{} at 6173 \AA{} with an effective free-spectral range (FSR) of 6 \AA{}. A blocking filter of 2.5 \AA{}
is introduced to restrict the FP channel of the desired wavelength. For the Ca II 8542 \AA{} line we use only one etalon, with 175 m\AA{} FWHM. 
As explained earlier, polarimeter consists of a two LCVRs and a linear polarizer (Glan-Thompson polarizer).

The optical setup of the spectropolarimeter is shown in Figure 14. F$\#$12 beam from the telescope and its re-imaging optics is collimated using a lens (L1) of 
focal length 500 mm. The FPs are placed in this collimated beam in order to reduce the wavelength shift produced by the finite angle of incidence arising from the 
FOV. Collimated output through the FPs is imaged using a lens (L2) of focal length 750 mm, which forms a plate scale of $0.145^{\prime\prime}$pixel$^{-1}$ at the 
CCD plane. The polarimeter is placed in this converging beam after the lens L2, to accommodate smaller Glan-Thompson polarizer. The acceptance angles of the LCVRs 
are large enough to work with resultant F$\#$18 beam. Since the z-cut etalons with a small angle of incidence produce negligible polarization effects, we have not 
separately considered the effect of the etalons in polarization measurements. The fast axes of the first and second LCVRs are kept at $0^{\circ}$ and $45^{\circ}$ 
with respect to the polarization axis of the linear polarizer as described in section 2. The temperatures of the LCVRs are kept at $28^{\circ}C$.

For the present set of observations, we are modulating polarization first and wavelength later to minimize the seeing influence. The response time of LCVRs 
(for the change of one polarization state to other) is 22 ms. Change in wavelength position requires 100 ms and 200 ms for a spectral 
sampling of 15 m\AA{}, and 30 m\AA{}, respectively as the tuning speed of the FPs is nearly 1000 Vs$^{-1}$.

% Presently the observation at each wavelength position is done by accumulating images for all the modulation steps by changing LCVR voltages, and then move 
% to the next wavelength step.
We capture 20 images for each polarizaton state to bulid-up the signal. Overall time taken for one 
modulation cycle (i.e. for obtaining IQUV at one wavelength poistion) is  around 8 seconds considering an exposure time of 60 ms (at 6173 \AA{}). 
Depending on the number of wavelength position the time cadance of the vector magnetogram varies; for \textit{e.g.},  the cadance varies from 40 seconds to
216 seconds for 5 to 27 wavelength positions, respectively. The number of wavelength points which determine the time cadance can be selected as required 
by the scientific objectives.
% SDO/HMI scans only 6 wavelength positions on the line profile, to retrieve the vector magnetic field \citep{Borrero2011a}, which is 
% used for studying the evolution of large scale magnetic field in the active regions. 
A cadance of more than a minute is sufficent enough to study the evolution of active regions and energy build-up due to the foot point motions 
\citep[and references therein]{Wiegelmann2012}. Magnetograms (by tuning the filter to a single wavelength position) can be obtained in 10 seconds cadance, 
which could be used for studying the magnetic field evolution of small scale moving magnetic 
features seen around the sunspot \citep[and references therein]{Ma2015}. A six wavelength step scan, which takes around 50 seconds can reproduce the line 
profile as seen in SDO/HMI, suitable for the vector magnetic field retrieval , as long as the horizontal speed of the moving feature is  below 3 kms$^{-1}$.

For the initial tests, we have scanned the spectral profile of 6173 \AA{} line with a step of 15 m\AA{} and 30 m\AA{}, for a total of 27 and 20 positions, 
respectively in longitudinal and vector modes. The number of wavelength positions could be considerably reduced by an optimization after inverting the profiles, 
which will be carried out after further analysis. We obtained observations in longitudinal and vector modes described in Section 2.1. 

 \begin{figure}   
   \centerline{\hspace*{0.015\textwidth}
               \includegraphics[width=0.5\textwidth,clip=]{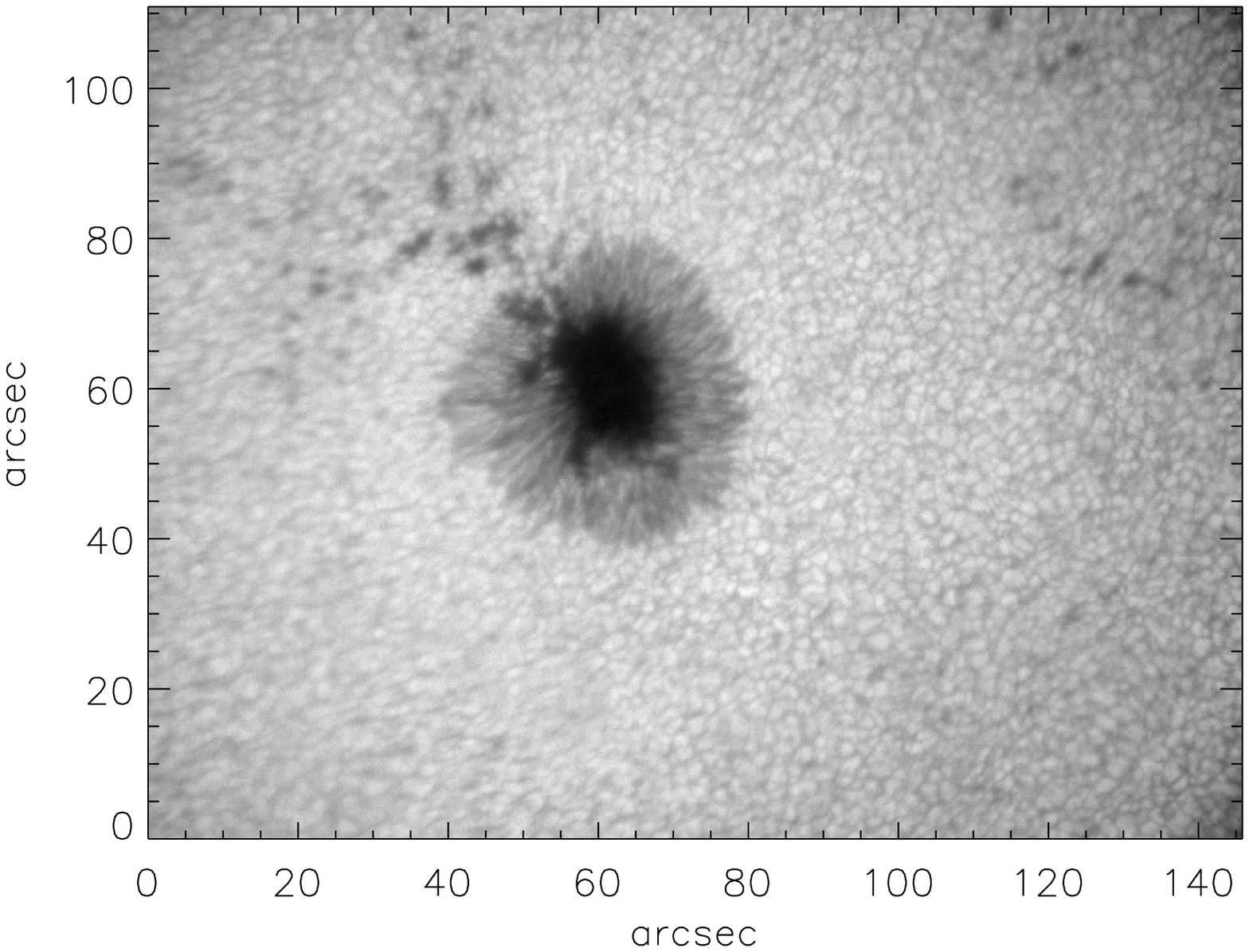}
               \hspace*{-0.03\textwidth}
               \includegraphics[width=0.5\textwidth,clip=]{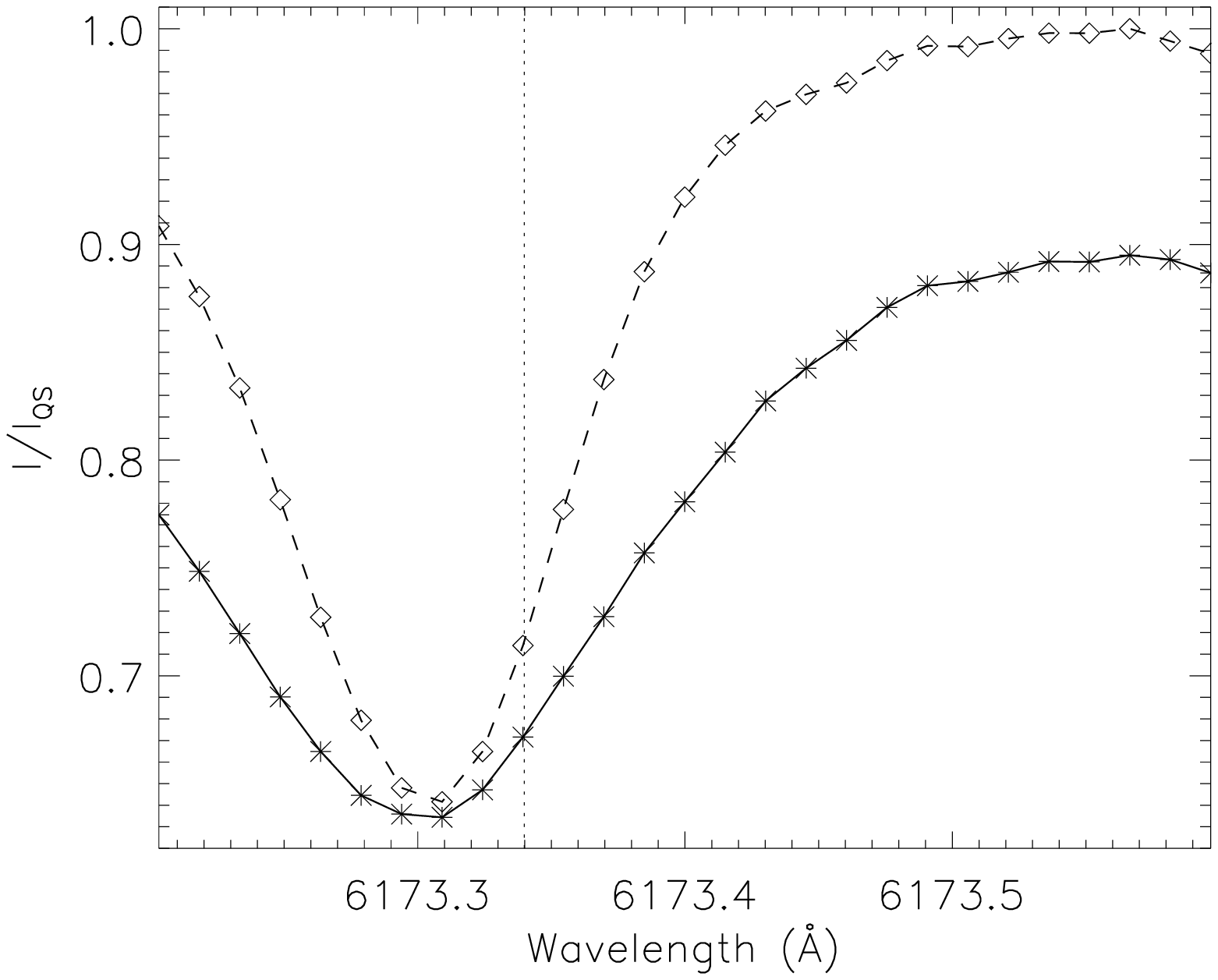}
              }
     \vspace{-0.35\textwidth}   % Shift close to the panel top 
     \centerline{\Large \bf     % Includes the labels (here needs the color 
                                %   package, see beginning of this file)
      \hspace{0.0 \textwidth}  %\color{white}{(a)}
      \hspace{0.5\textwidth}  %\color{white}{(b)}
         \hfill}
     \vspace{0.31\textwidth}    % Shift back to the panel bottom 
          
  \centerline{\hspace*{0.015\textwidth}
             \includegraphics[width=0.52\textwidth,clip=]{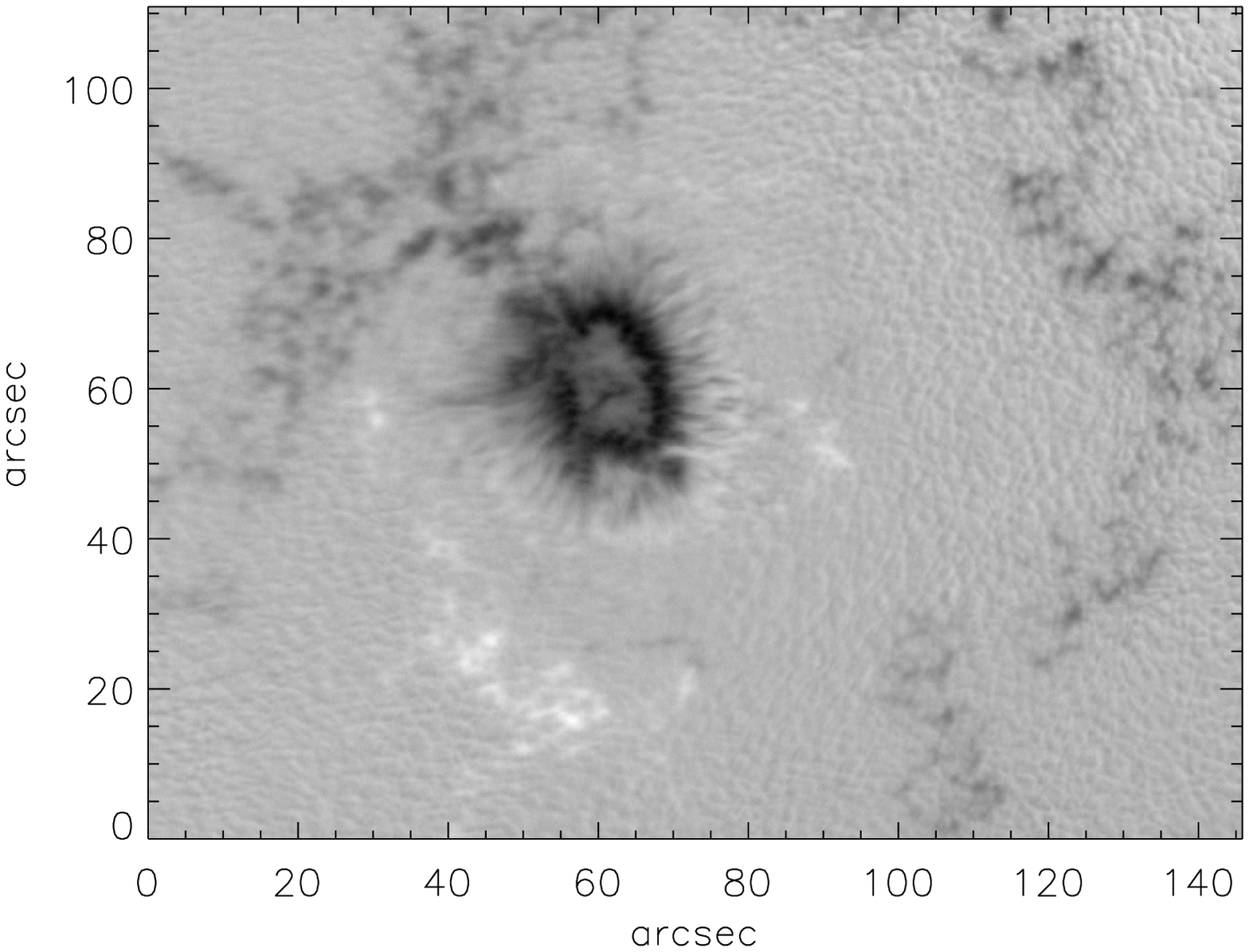}
             \hspace*{-0.03\textwidth}
              \includegraphics[width=0.52\textwidth,clip=]{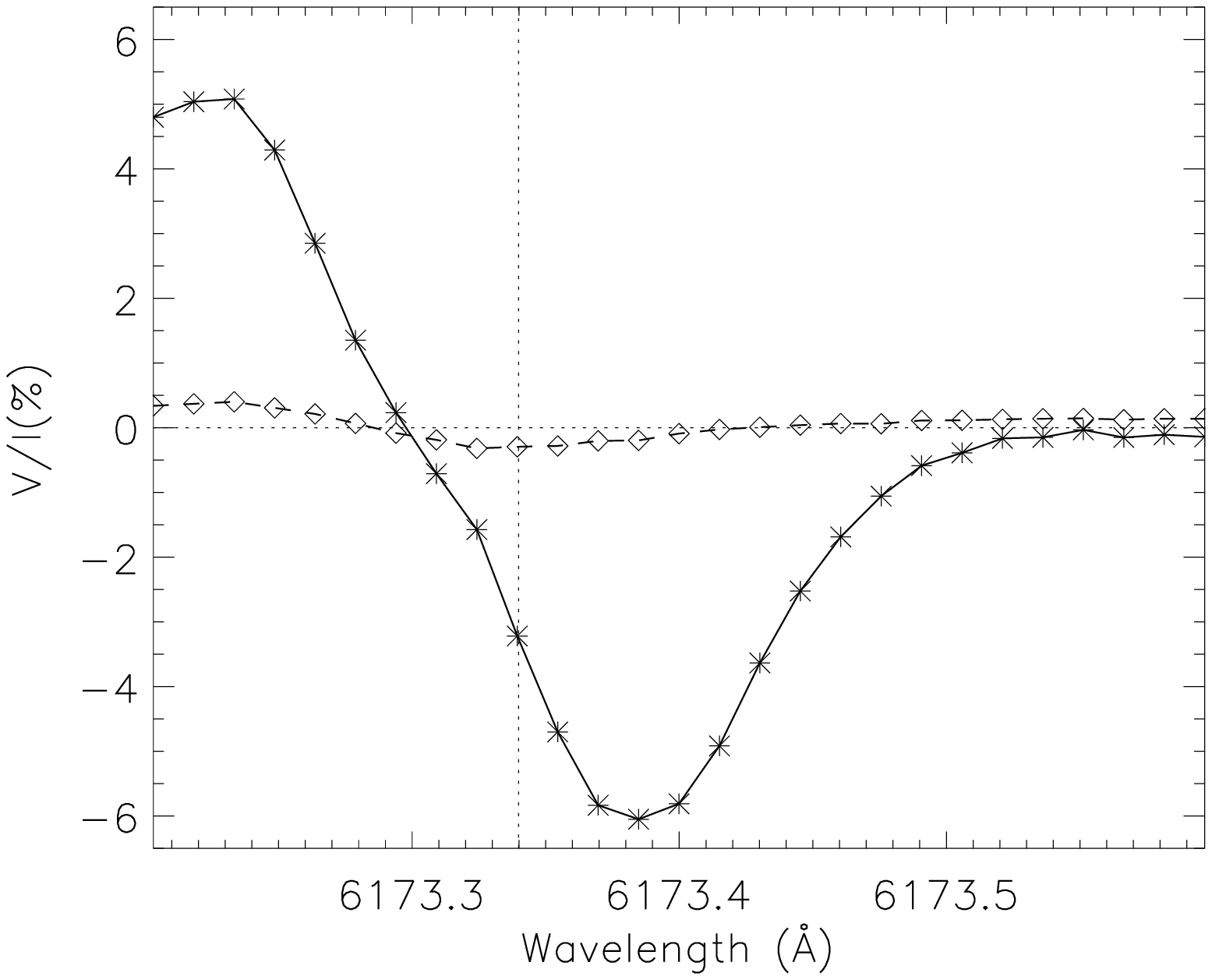}
             }
    \vspace{-0.35\textwidth}   % Shift close to the panel top 
    \centerline{\Large \bf     % Includes the labels (here needs the color package)
     \hspace{0.0 \textwidth} %\color{white}{}
     \hspace{0.415\textwidth}  %\color{white}{}
        \hfill}
    \vspace{0.31\textwidth}    % Shift back to the panel bottom 
             
\caption{LOS mode of observations of the active region NOAA AR 12436 observed on 24th October 2015 between 4:00 UT and 5:30 UT using MAST polarimeter. In the 
first row, the left Figure shows one of the mean intensity images whereas the right Figure is for the corresponding mean Stokes I profiles. The Stokes I profile 
is deduced separately for the magnetic (solid line) and for the non-magnetic (dashed lines) regions whereas in the second row, left Figure displays the mean Stokes 
V image for a wavelength position at $+75$ m\AA{} from line center and right plot indicates the mean Stokes V profile for both the magnetic and non-magnetic regions. 
Because of the limitation of the voltage tuning of both the etalons in tandem, line profile shifted more toward blue side.}
   %\label{fig_16}
   \end{figure}	
   
\subsection{Longitudinal mode}
The observations described in this section were obtained for a sunspot in the active region NOAA AR 12436 taken on 24th October 2015 between 4:00 UT and 5:30 UT, 
when the seeing was moderate. The active region was slightly away from the disk center located at N09 and W20. FP etalons of the 
narrow-band imager were sequentially tuned to 27 positions on the 6173\AA{} line, with 15 m\AA{} wavelength spacing. A pair of two images in left- and right- circular polarizations (LCP \& RCP) was obtained by applying 
appropriate voltages (listed in Table 5) to the LCVR 1 and 2. For these measurements, the voltage of the LCVR1 is changed alternately, whereas the LCVR2 is kept 
constant to provide a retardance of 1$\lambda$. For each wavelength position, 20 pairs of LCP and RCP images were obtained with an exposure time of 65 ms 
to increase the signal to noise ratio (SNR). Figure 16 shows the results of the above observation. The top left Figure shows one of the mean intensity image whereas 
the right Figure is for the corresponding mean Stokes I profiles. The mean Stokes I profile is deduced separately for both the magnetic (solid line, where the V 
signal is more than $10^{-3}$) and for the non-magnetic (dashed line) regions. The broadening of the profile due to the magnetic field is evident in these plots. 
The bottom left Figure displays the mean Stokes V image for a wavelength position at $+75$ m\AA{} from line center whereas the right plot indicates the mean 
Stokes V profile for both the magnetic and non-magnetic regions.

\subsection{Comparison of Stokes V images from SDO/HMI and USO/MAST}
We also carried out a comparison of our results with the magnetograms availed from \textit{Helioseismic Magnetic Imager} (HMI) instrument 
\citep{Scherrer2012, Schou2012} onboard the  \textit{Solar Dynamics Observatory} (SDO) \citep{Pesnell2012}. For comparison, the images from MAST and HMI were 
taken at around the same time (04:42 UT on 24th October 2015). The comparison was possible as the spectral line used by both the instruments is same, even though 
the spectral resolution of each instrument differs. SDO/HMI images are available for both the continuum and the line-of-sight (LOS) magnetic field, and also for 
the Stokes I, Q, U and V parameters as for right, left, circular and linear polarization images. For our present comparison, we constrained 
only to SDO/HMI LOS magnetograms and Stokes V images, since our polarization measurements need further instrumental polarization corrections.
Even though the stokes V profiles could also get contaminated by the instrumental polarization, for a preliminary comparison the Stokes V images are suitable 
as the cross-talk from the linear polarization mostly introduce only a bias. 
\newline
The images obtained from the SDO/HMI is resized, and the USO/MAST image is registered with respect to SDO/HMI image. 
The right and left images in the top panel of Figure 17 shows the cropped continuum images taken with SDO/HMI and USO/MAST instruments,  respectively. The images 
are cropped in such a way that it include the sunspot and a part of the nearby area. The middle panel show the Stokes V images of SDO/HMI (left) and USO/MAST 
polarimeter (right), respectively for a selected wavelength position $+75$ m\AA{} (from the line center). It is evident from Figures that most of the magnetic 
features in the Stokes V map of SDO/HMI matches well with the Stokes V map of USO/MAST polarimeter. The advantage of the space-based observation is clearly 
visible in the SDO/HMI images as evident from the low background features.

Unlike SDO/HMI observations, the USO/MAST images are affected by the atmospheric seeing during the image acquisition which results in a considerable 
I$\rightarrow$V cross-talk \citep{2003isp..book.....D, Lites1987} thus a higher background noise.
This I$\rightarrow$V cross-talk is evident from the granulation pattern in Stokes V images of USO/MAST.

The bottom Figures are for the Stokes V signal obtained at a wavelength position $+75$ m\AA{} (from the line center) plotted against the SDO/HMI LOS magnetic 
field strength. The plot shown in the left panel is for the USO/MAST Stokes V whereas the right is for the SDO/HMI Stokes V at a close by wavelength position
against the SDO/HMI LOS magnetic field strength. The region shown with the red contain points mostly from the umbra, where the linearity between the 
Stokes V amplitude and the magnetic field strength doesn't hold. As it is evident in Figure 16, other than the scatter which could be partly due to 
the seeing related I$\rightarrow$V cross-talk, the trend matches closely. 

We also found a factor of around 2 in the Stokes V signal between the SDO/HMI and the USO/MAST images. 
This can be explained by the influence of finite width of the narrow band filter in scanning the line profile and the difference in the overall 
instrumental profiles used in the HMI and MAST imager. The finite width of the filter results in a 
convolution of the spectral line profile, which introduces an apparant increase in the width and decrease in the depth of the line profile. This effectively 
reduces the amplitude of the Stokes V signal. In order to check this we have computed, synthetic line profiles using M-E inversion code, taking realistic solar 
atmospheric parameters. Convolution of the Stokes V parameter with the filter profile of 95 m\AA{} FWHM shows a reduction in the peak of Stokes V amplitude by 
a similar factor, \textit{i.e.}, 2. This effect can be taken care while inverting the Stokes profiles; \textit{i.e.}, convolution of the synthetic profile with filter profile is 
carried out before fitting that with observed Stokes profiles. Other than the above differences, the overall comparison between the SDO/HMI and the USO/MAST 
Stokes V measurements provide the confidence in our measurements.

 \begin{figure}   
   \centerline{\hspace*{0.015\textwidth}
               \includegraphics[width=0.5\textwidth,clip=]{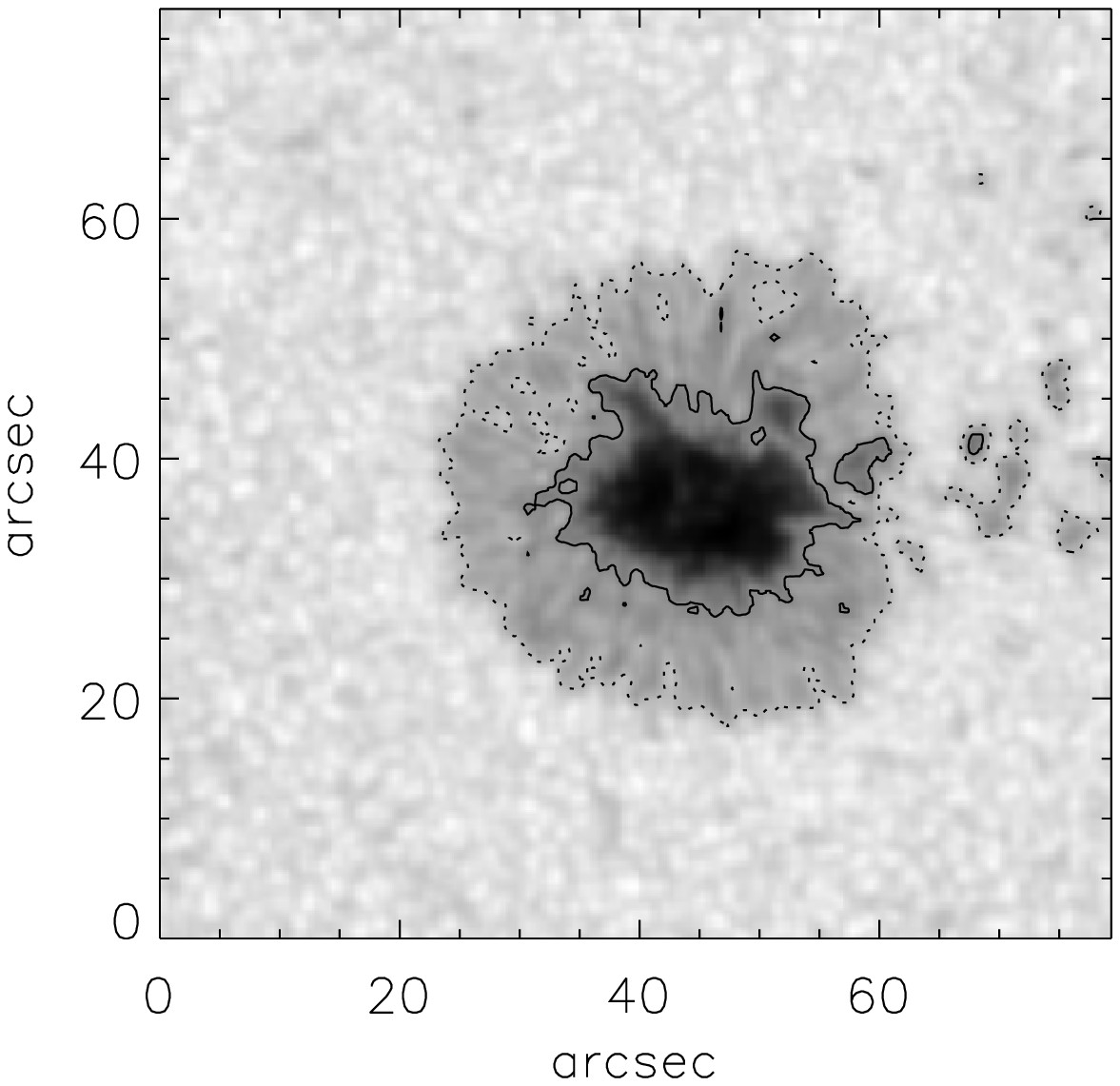}
               \hspace*{-0.03\textwidth}
               \includegraphics[width=0.5\textwidth,clip=]{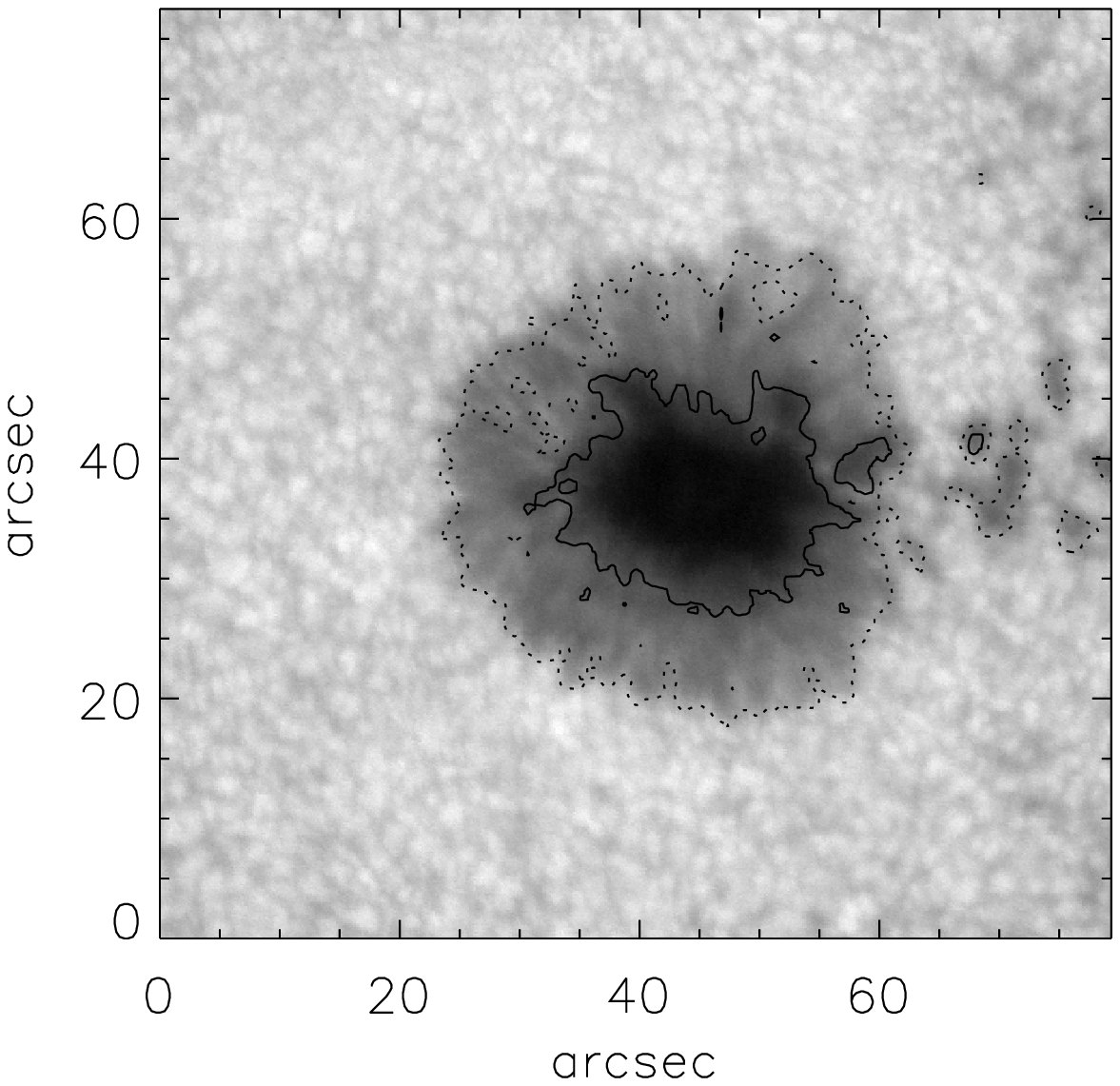}
              }
     \vspace{-0.35\textwidth}   % Shift close to the panel top 
     \centerline{\Large \bf     % Includes the labels (here needs the color 
                                %   package, see beginning of this file)
      \hspace{0.0 \textwidth}  %\color{white}{(a)}
      \hspace{0.5\textwidth}  %\color{white}{(b)}
         \hfill}
     \vspace{0.31\textwidth}    % Shift back to the panel bottom 
          
  \centerline{\hspace*{0.015\textwidth}
             \includegraphics[width=0.5\textwidth,clip=]{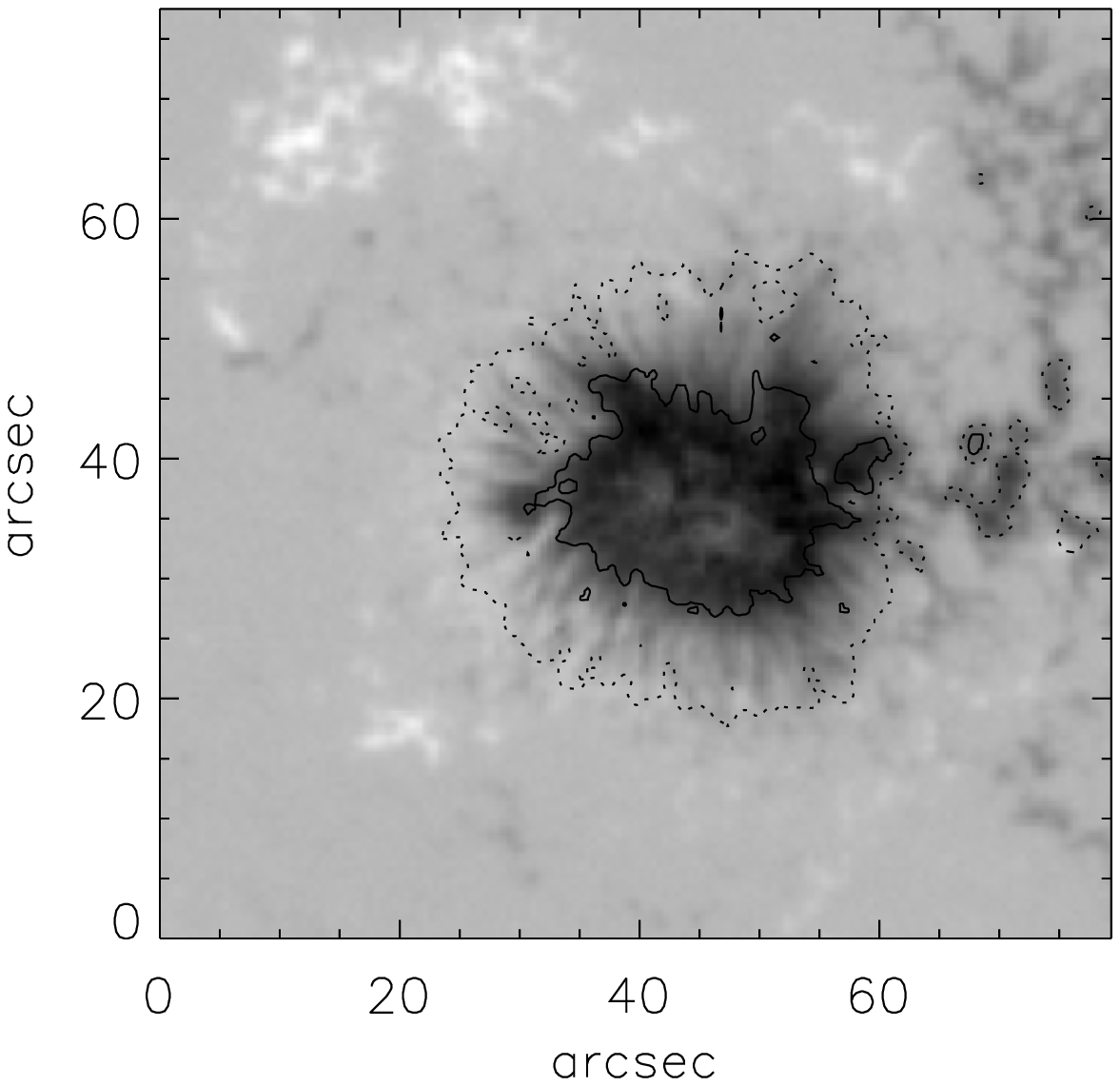}
             \hspace*{-0.03\textwidth}
              \includegraphics[width=0.5\textwidth,clip=]{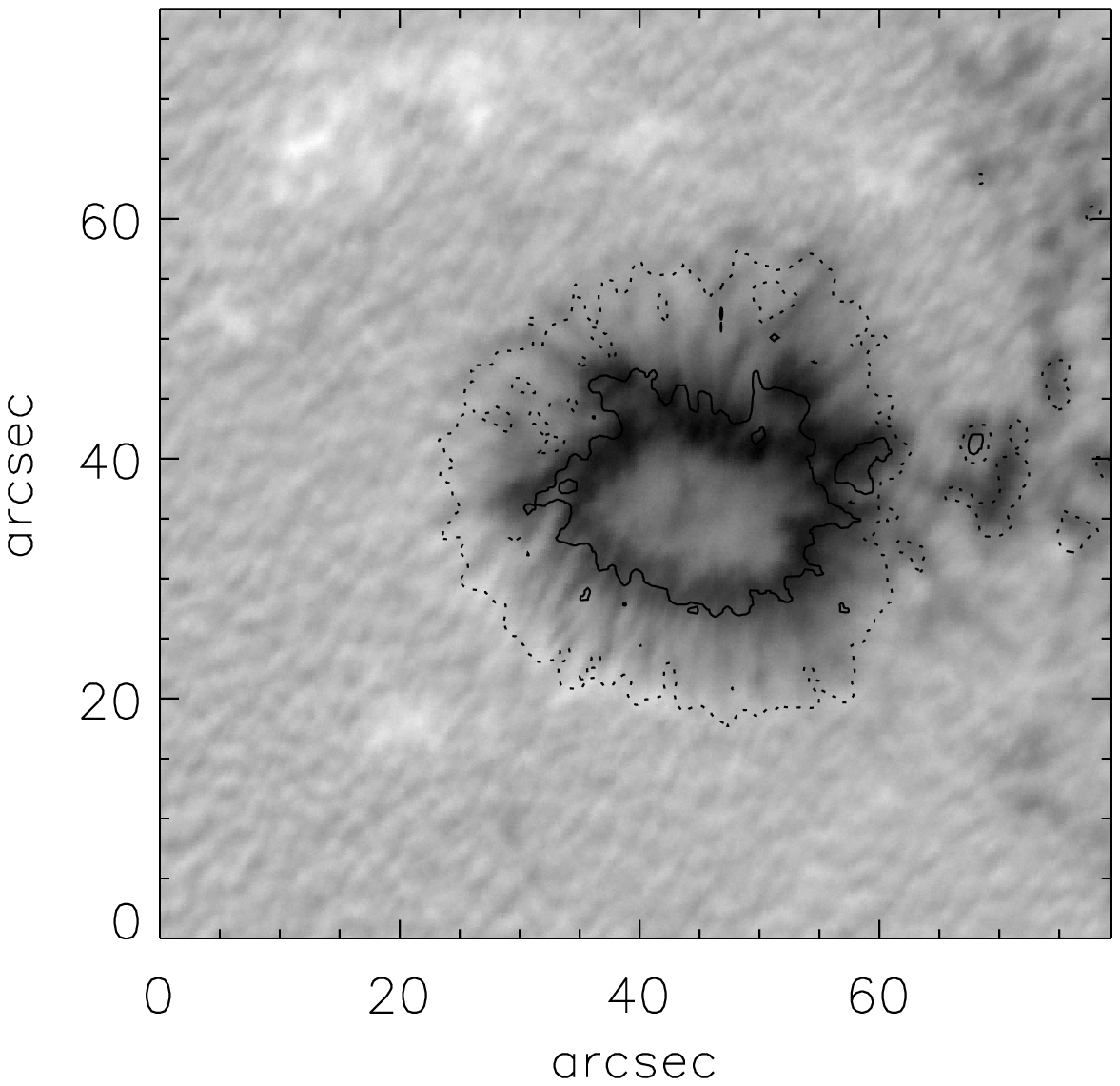}
             }
    \vspace{-0.35\textwidth}   % Shift close to the panel top 

 \vspace{0.31\textwidth}    % Shift back to the panel bottom 
          
  \centerline{\hspace*{0.015\textwidth}
             \includegraphics[width=0.5\textwidth,clip=]{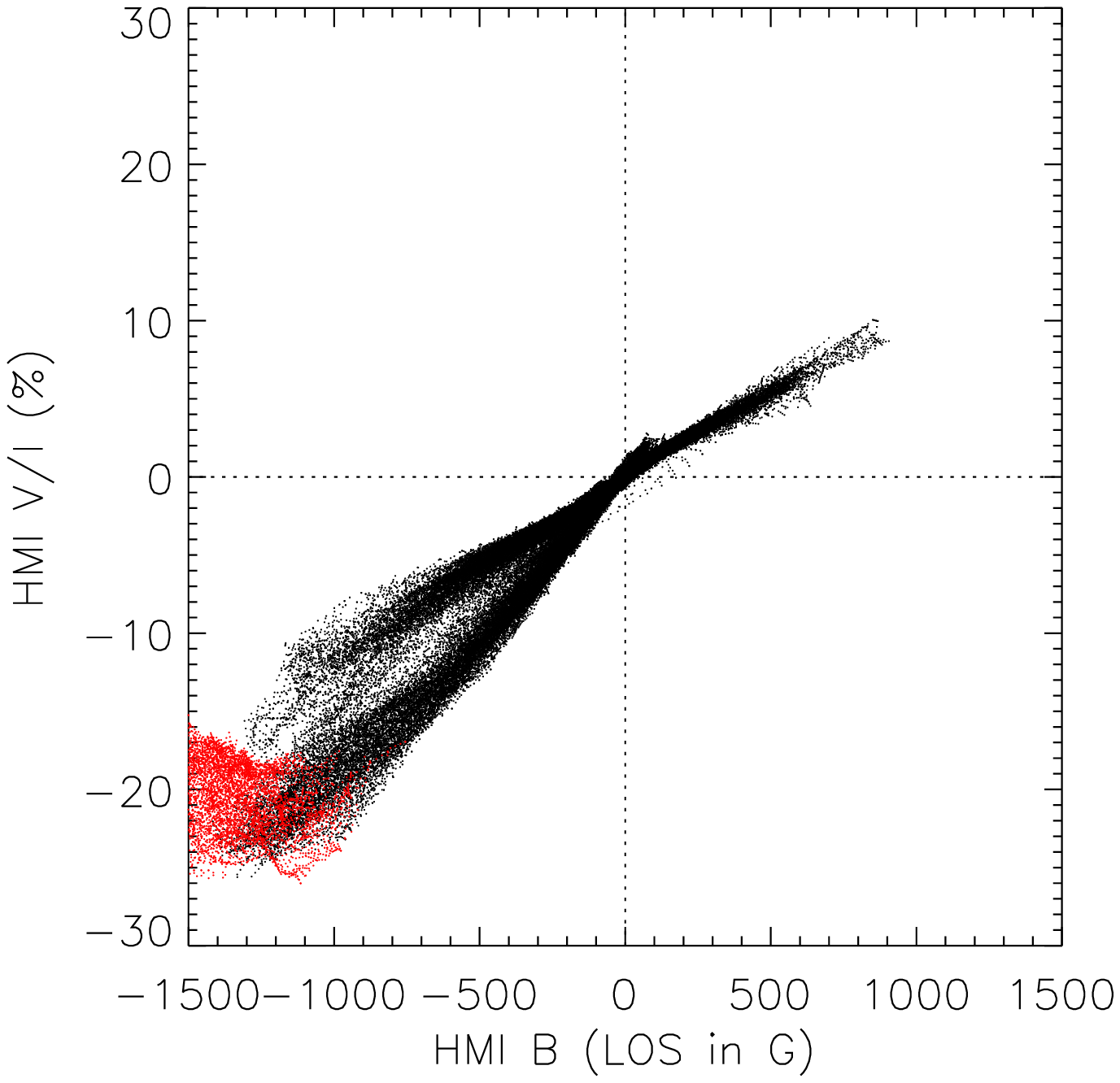}
            \hspace*{-0.03\textwidth}
              \includegraphics[width=0.5\textwidth,clip=]{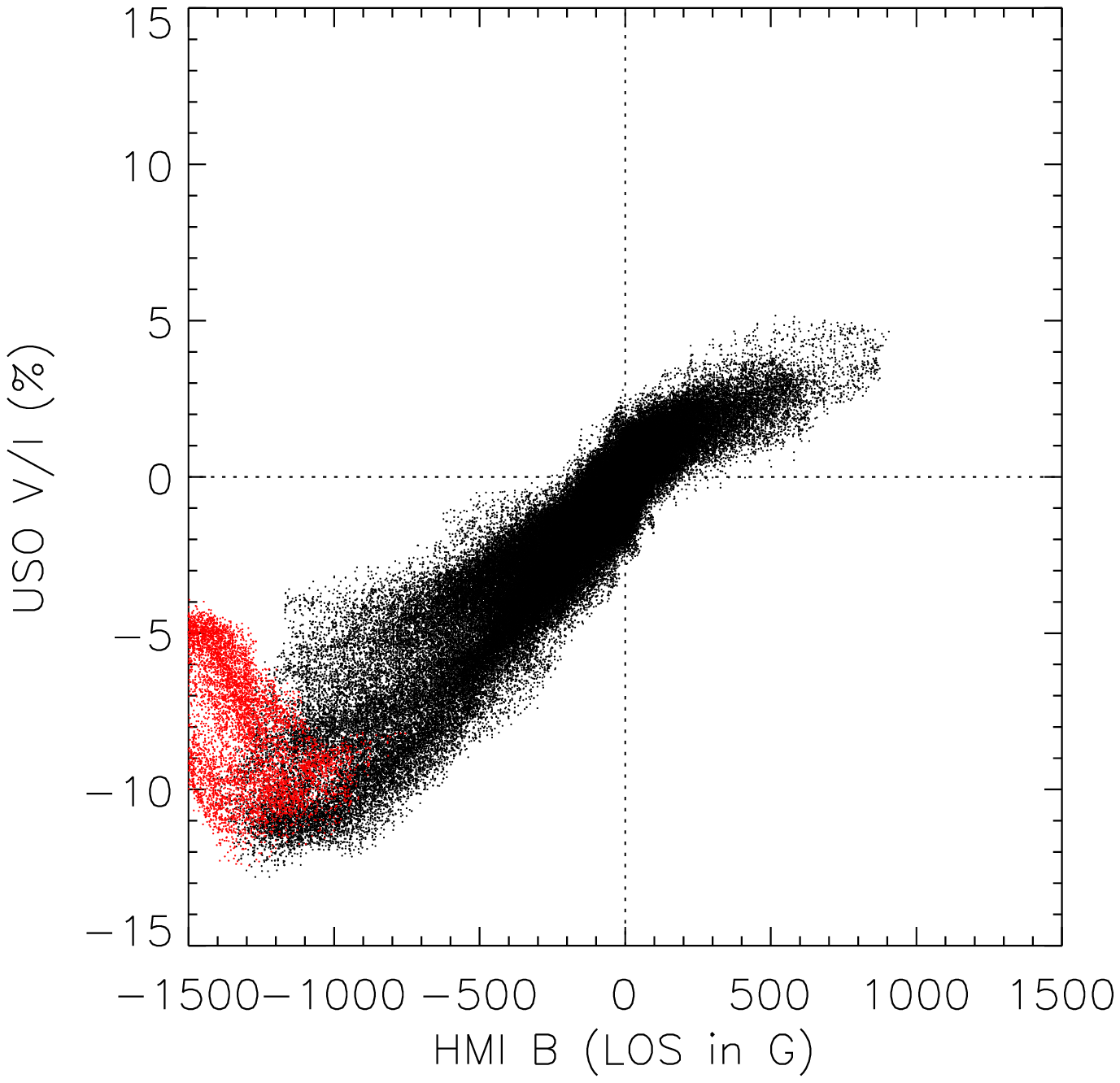}
             }
    \vspace{-0.35\textwidth}   % Shift close to the panel top 
    \centerline{\Large \bf     % Includes the labels (here needs the color package)
     \hspace{0.0 \textwidth} %\color{white}{}
     \hspace{0.415\textwidth}  %\color{white}{}
        \hfill}
    \vspace{0.31\textwidth}    % Shift back to the panel bottom 
             
\caption{Top row: Stokes I image taken by SDO/HMI (left) and USO/MAST (right); Middle row: Stokes V image at wavelength position $+75$ m\AA{} from line 
center taken by SDO/HMI (left) and USO/MAST (right); bottom row: scatter plot made between Stokes V of SDO/HMI and longitudinal magnetic field of SDO/HMI (left) 
and scatter plot made between Stokes V of USO/MAST and longitudinal magnetic field of SDO/HMI (right).}
%\label{fig_17}
 \end{figure}
    \subsection{Vector mode}
By operating the imaging polarimeter in vector mode, we have carried out observations on $19^{th}$ December 2015.
The active region NOAA AR12470 (N15, W07) was observed during the period 06:00 UT and 07:30 UT. Four images were obtained sequentially by applying appropriate 
voltages (listed in Table 4) to the LCVRs. From the observed intensity images Stokes, I, Q, U and V images were computed for 
 each wavelength position. Similar to the longitudinal mode the images in vector mode acquired by scanning  the line profile with 
15 m\AA{} spacing and at 27 wavelength positions. Figure 17 shows the images, top left and right panels shows the  
Stokes I, and Q, the bottom left and right panels show the U and V images, respectively. The Stokes U, Q, and V 
images are shown for a wavelength position at $+75$ m\AA{} from line center on the line profile. The bottom panel shows 
respective Stokes profiles for the single point marked by a star in the  Stokes I image. A thorough analysis and 
demodulation of the linear polarization measurement requires the knowledge of the instrumental polarization. 
For this purpose, we are currently introducing a large (50 cm) sheet linear polarizer which can be rotated, in front 
of the primary mirror. This will enable us to characterize the telescope polarization before extracting the Stokes 
information from these measurements.

\begin{figure}   
   \centerline{\hspace*{0.015\textwidth}
               \includegraphics[width=0.5\textwidth,clip=]{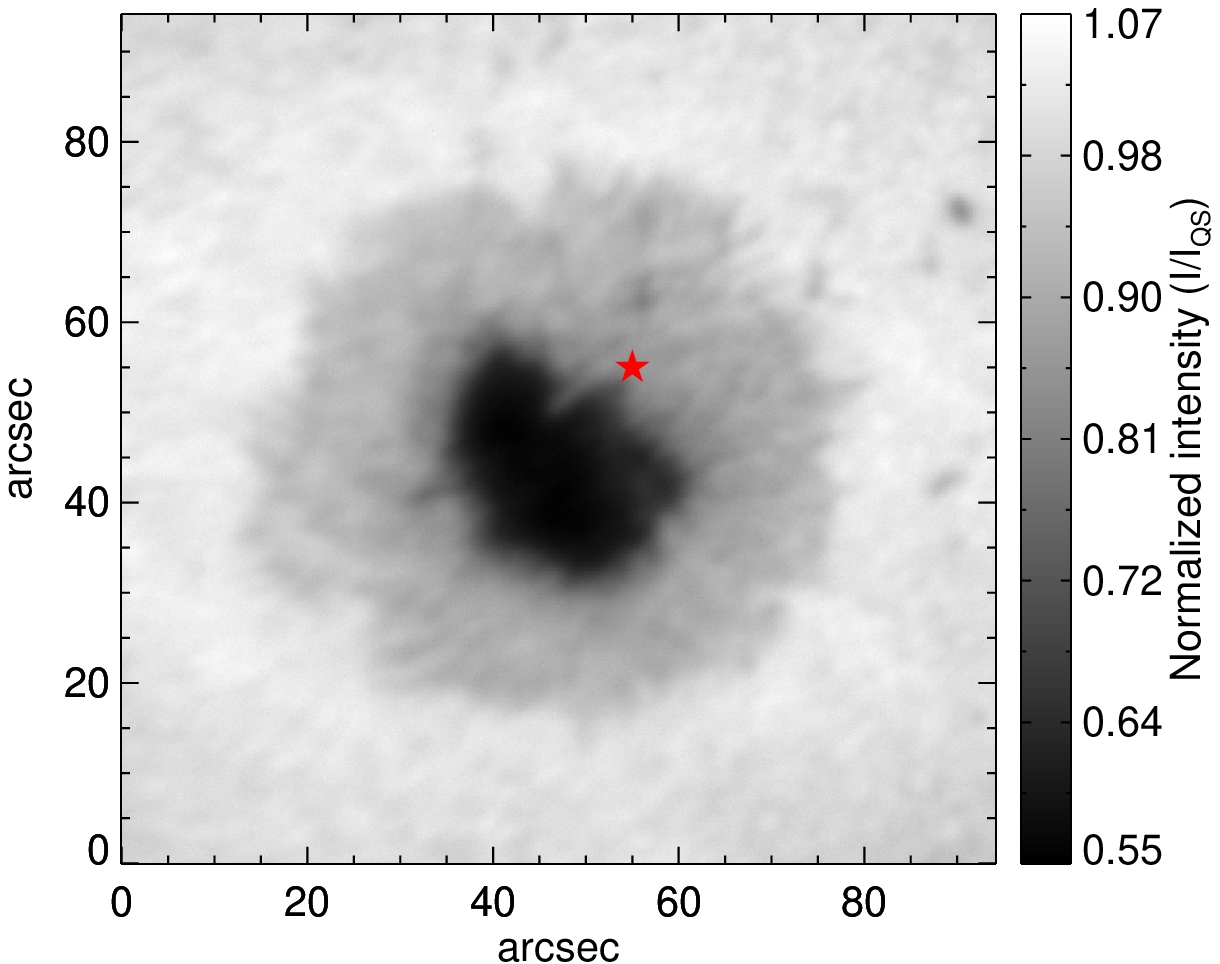}
               \hspace*{-0.03\textwidth}
               \includegraphics[width=0.5\textwidth,clip=]{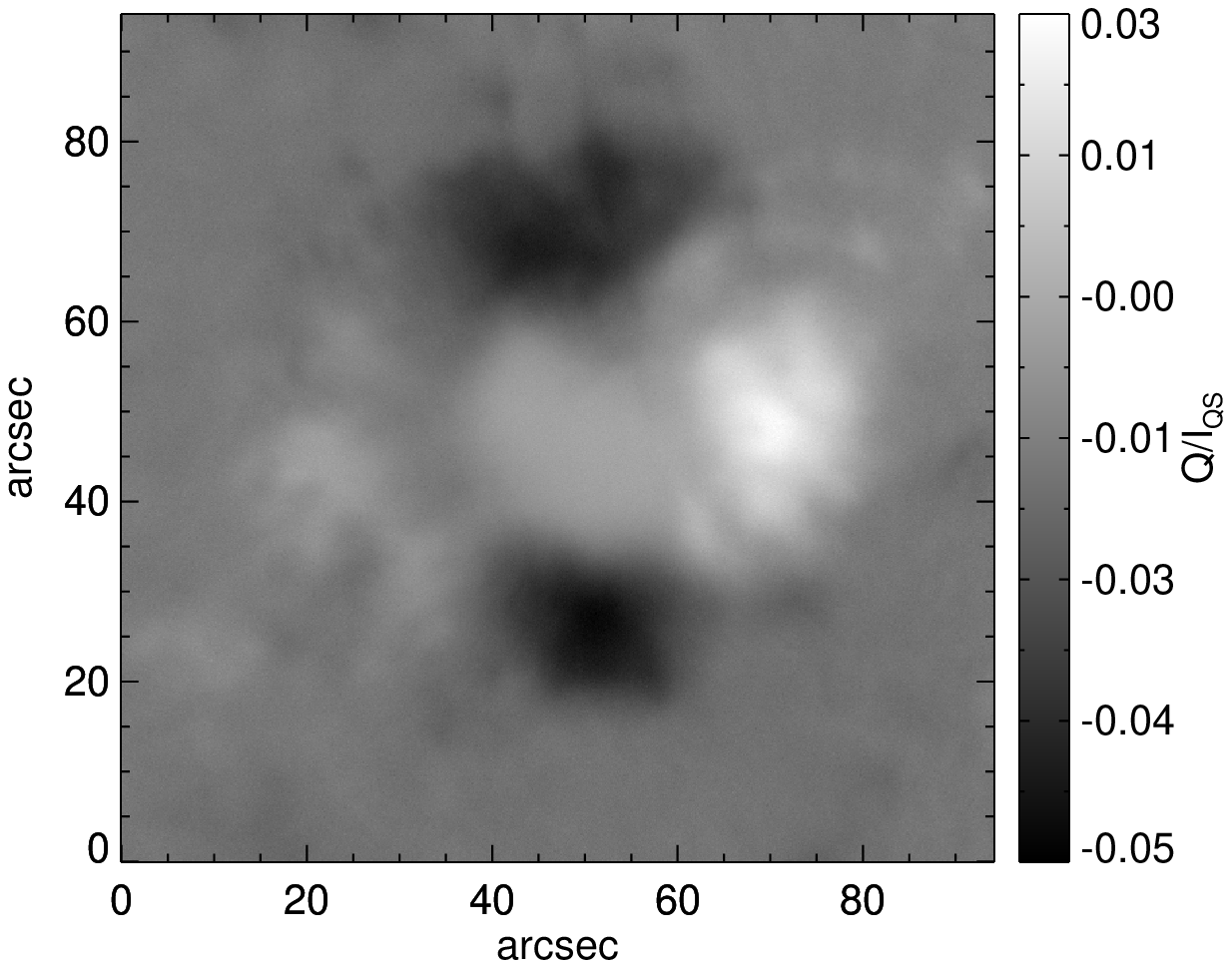}
              }
     \vspace{-0.35\textwidth}   % Shift close to the panel top 
     \centerline{\Large \bf     % Includes the labels (here needs the color 
                                %   package, see beginning of this file)
      \hspace{0.0 \textwidth}  %\color{white}{(a)}
      \hspace{0.415\textwidth}  %\color{white}{(b)}
         \hfill}
     \vspace{0.31\textwidth}    % Shift back to the panel bottom 
          
  \centerline{\hspace*{0.015\textwidth}
             \includegraphics[width=0.5\textwidth,clip=]{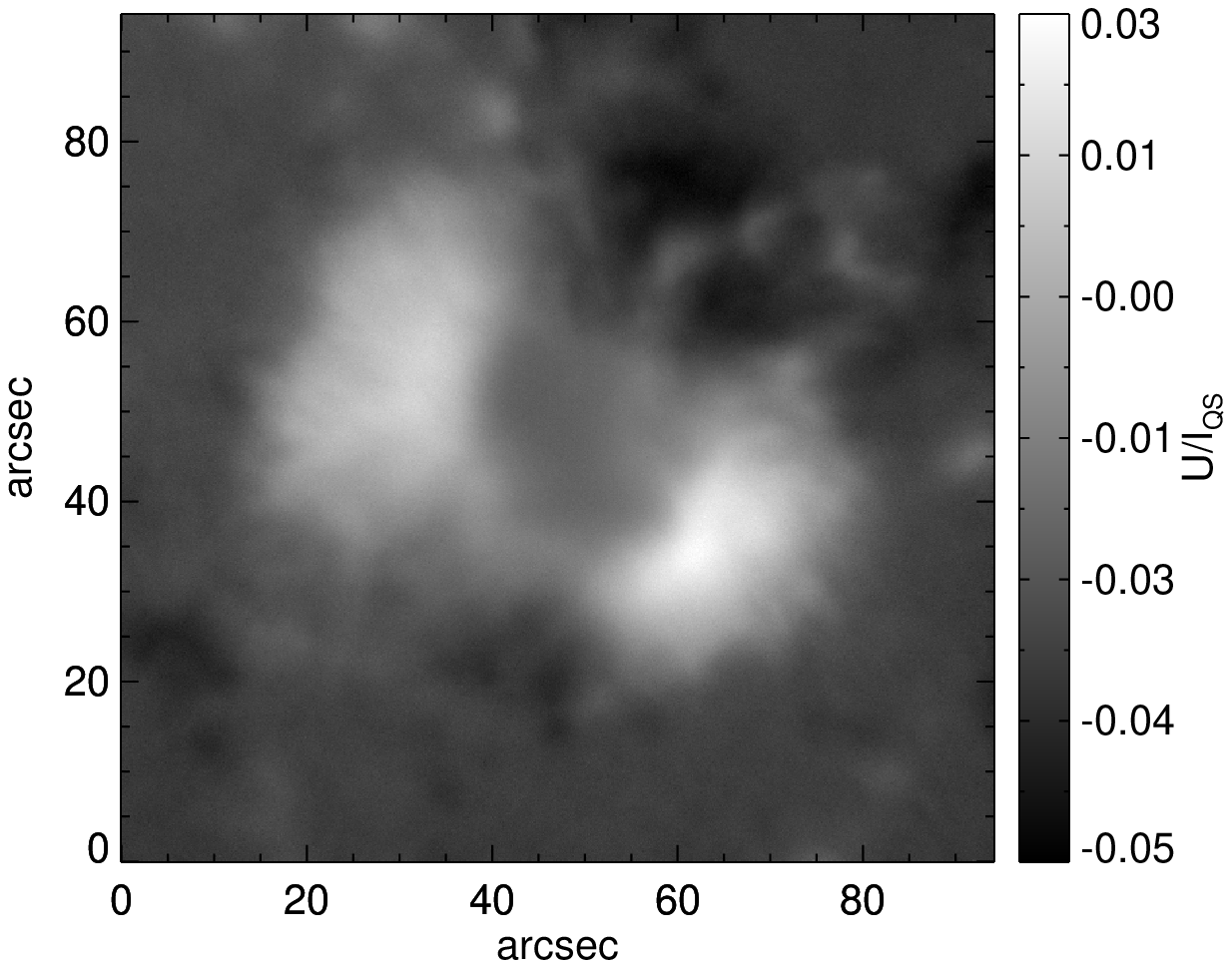}
             \hspace*{-0.03\textwidth}
              \includegraphics[width=0.5\textwidth,clip=]{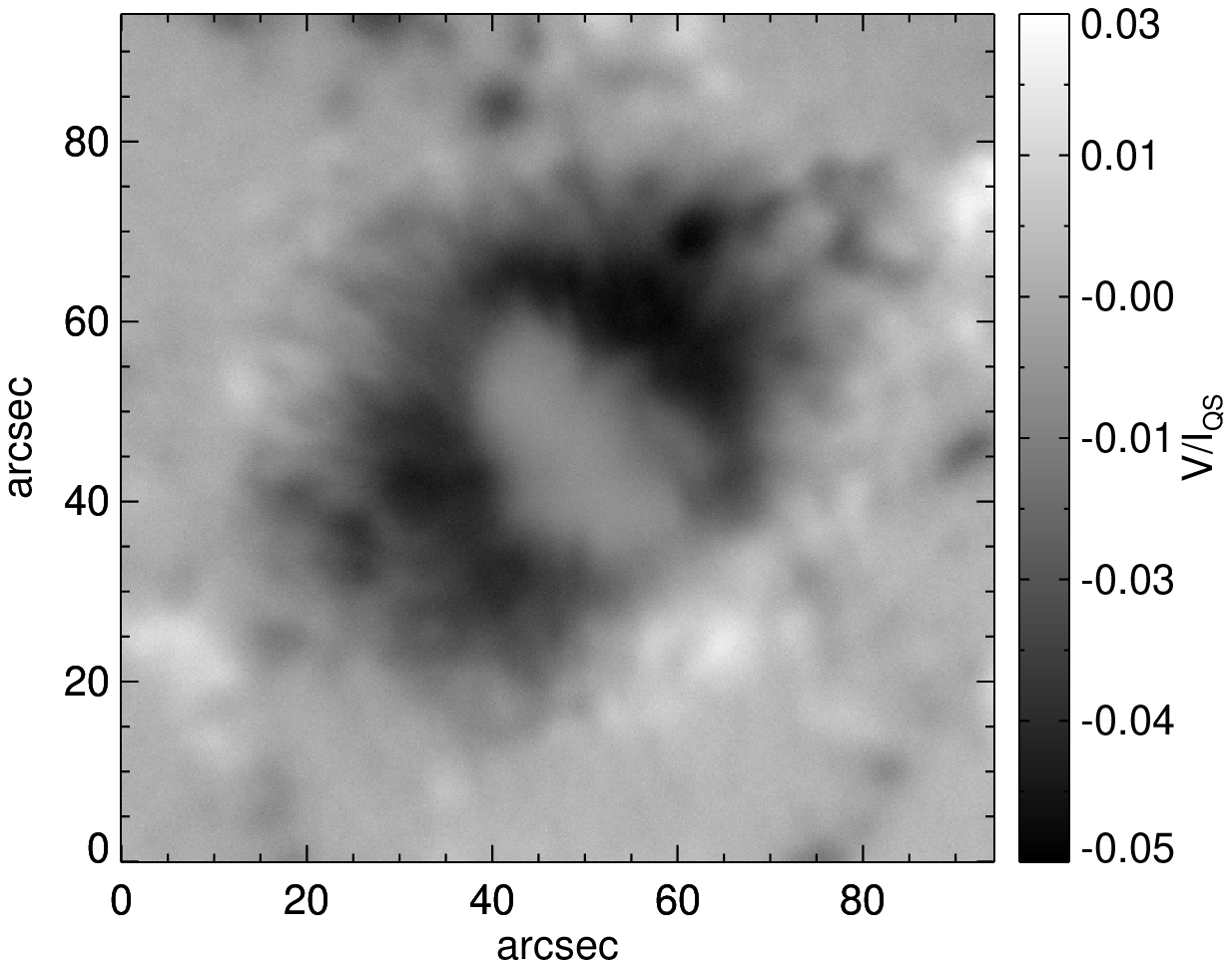}
             }
    \vspace{-0.35\textwidth}   % Shift close to the panel top 
    \centerline{\Large \bf     % Includes the labels (here needs the color package)
     \hspace{0.0 \textwidth} %\color{white}{}
     \hspace{0.415\textwidth}  %\color{white}{}
        \hfill}
    \vspace{0.31\textwidth}    % Shift back to the panel bottom 
             
\caption{Stokes images obtained using vector mode of operation for NOAA AR 12470 at 6173 \AA{} are shown here. Top panels show Stokes I (left), Q (right).
Bottom panels show Stokes U (left) and V (right).}
   %\label{fig_18}
   \end{figure}

\begin{figure}
\centerline{\includegraphics[width=1.0\textwidth,clip=]{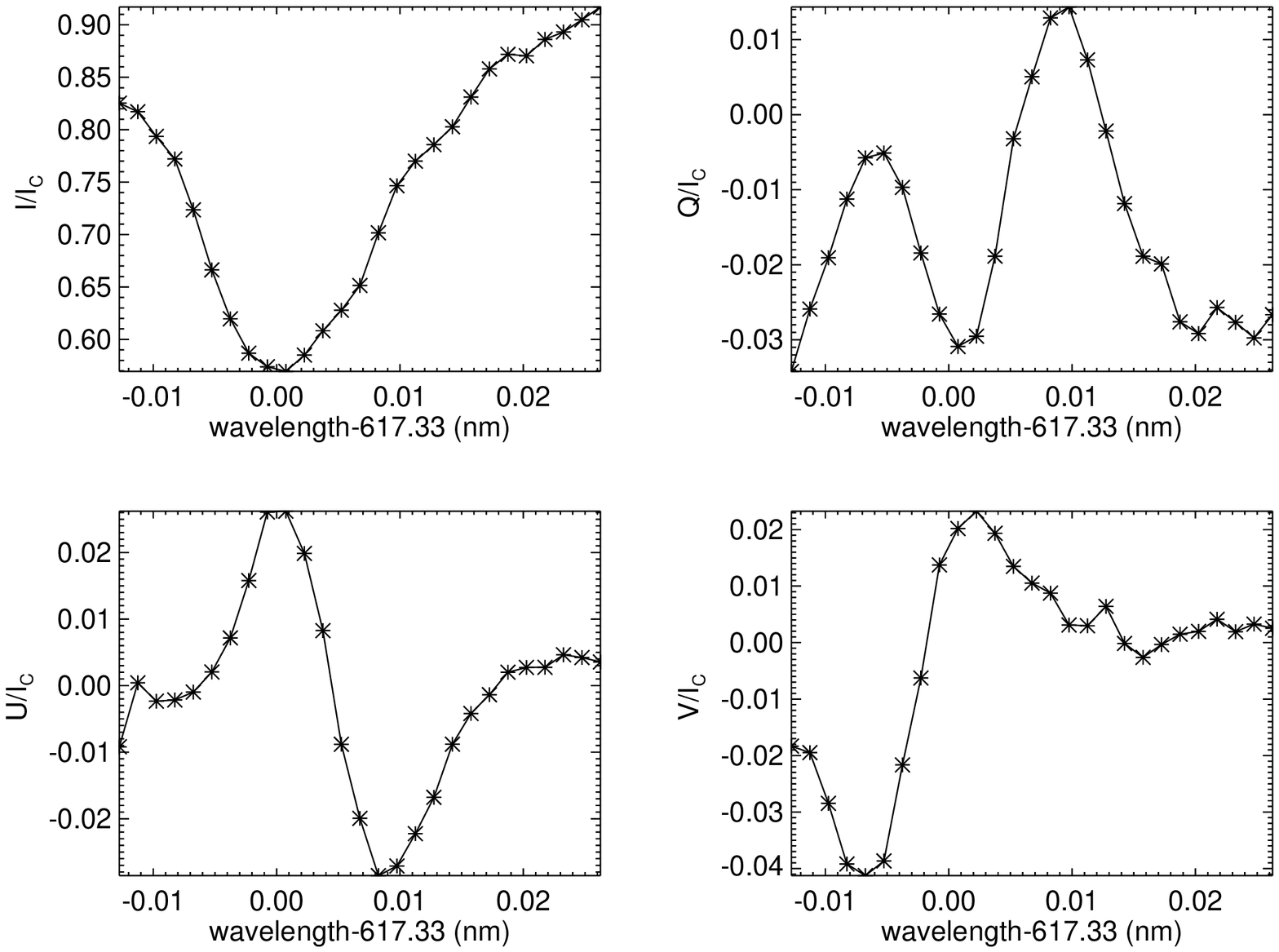}
           }
	\caption{Profiles of Stokes parameters I, Q, U, and V at a point in the penumbra of NOAA AR 12470 corresponding to the star mark in Figure 17.}
	 
	%\label{fig_19}
\end{figure}

\subsection{Observations on 16 April, 2016}
As evident in the Stokes I profile in the top panel of Figure 15, the starting point for the wavelength scanning has a limited coverage in the blue wing 
side of the Fe I 6173 \AA{} line profile. This was due to the limited tuning range of the Fabry-Perot filters, which was 
restricted due to the maximum voltage which could be applied to the etalons, and the operating temperatures of the etalons. 
As it is important to cover the entire wavelength range in order to obtain the continuum intensity at both sides of the line profile, 
we have carried out a re-tuning of the etalons to optimally cover the continuum at both the sides. Since the maximum 
allowed voltage which could be applied to the etalon is restricted, the re-tuning was done by changing the operating temperature. 
The re-tuning allowed us to start the line profile scan from a wavelength point further blue in the continuum. The following example of the Stokes V scan was 
carried after the re-tuning of the filters. These observations were taken on $16^{th}$ April 2016. 
Here the sunspot in the active region NOAA AR 12529 (N10, W38) was observed with the  polarimeter. The observations were obtained between 07:00 UT and 07:30 UT.
During data acquisition, the seeing was again moderate. Unlike the previous observations, instead of 27 wavelength positions, we increased the wavelength spacing 
to 30 m\AA{} which resulted in around 20 spectral positions on 6173 \AA{} line for the wavelength scan. Figure 19 shows the images for the wavelength 
position $+75$ m\AA{} from the line center and the corresponding mean Stokes V profile (right) for the entire FOV. In this case, the wavelength scan started 
well in the blue continuum, and the Stokes V signal also covers enough continuum wavelength points. 
 
 \begin{figure}   
   \centerline{\hspace*{0.015\textwidth}
               \includegraphics[width=0.5\textwidth,clip=]{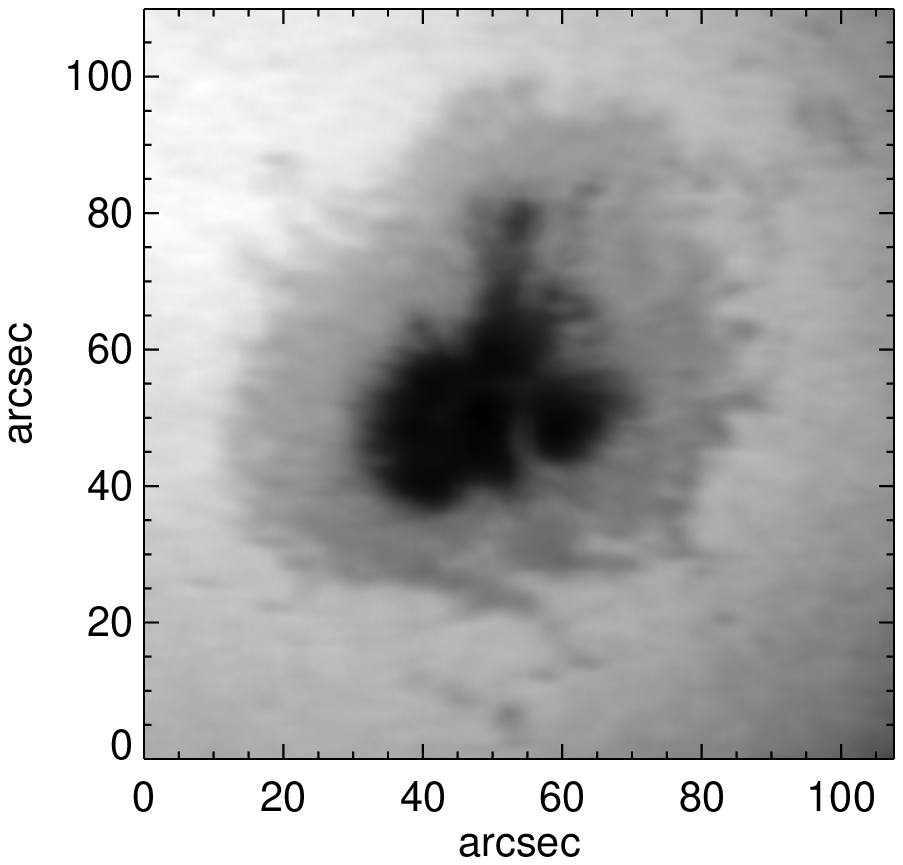}
               \hspace*{-0.03\textwidth}
               \includegraphics[width=0.5\textwidth,clip=]{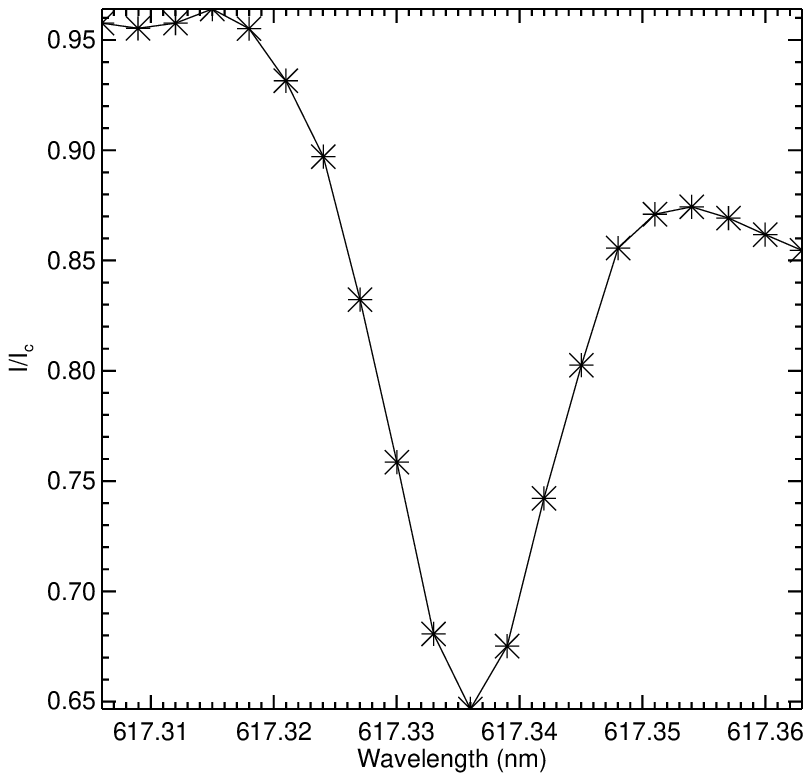}
              }
     \vspace{-0.35\textwidth}   % Shift close to the panel top 
     \centerline{\Large \bf     % Includes the labels (here needs the color 
                                %   package, see beginning of this file)
      \hspace{0.0 \textwidth}  %\color{white}{(a)}
      \hspace{0.415\textwidth}  %\color{white}{(b)}
         \hfill}
     \vspace{0.31\textwidth}    % Shift back to the panel bottom 
          
  \centerline{\hspace*{0.015\textwidth}
             \includegraphics[width=0.5\textwidth,clip=]{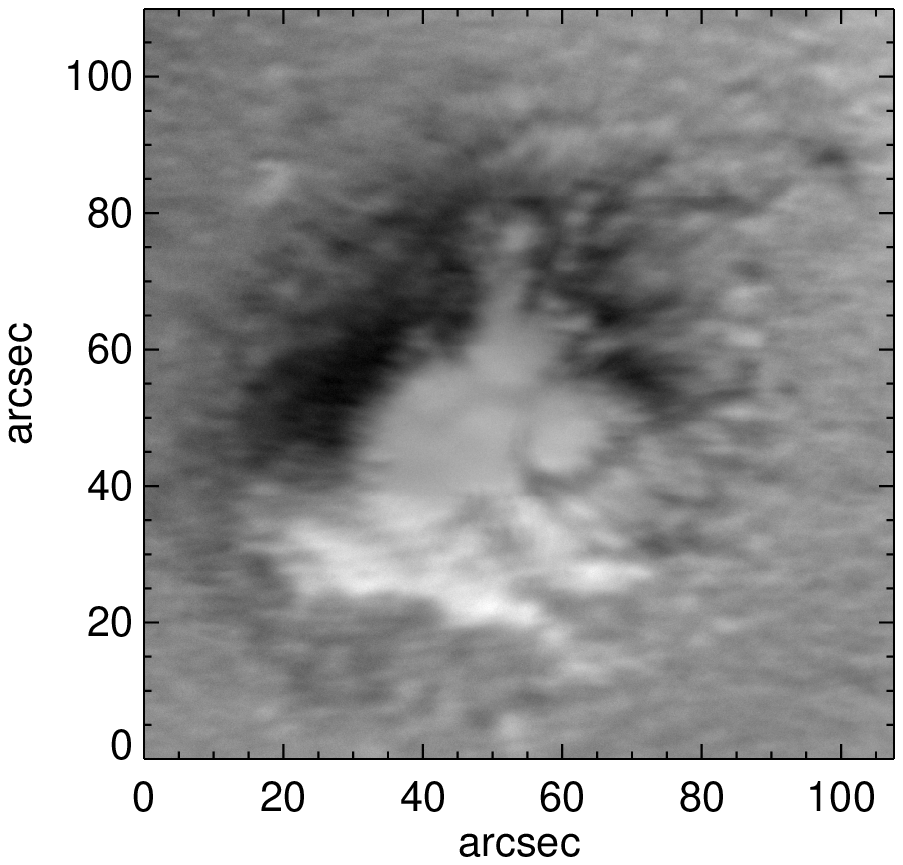}
             \hspace*{-0.03\textwidth}
              \includegraphics[width=0.5\textwidth,clip=]{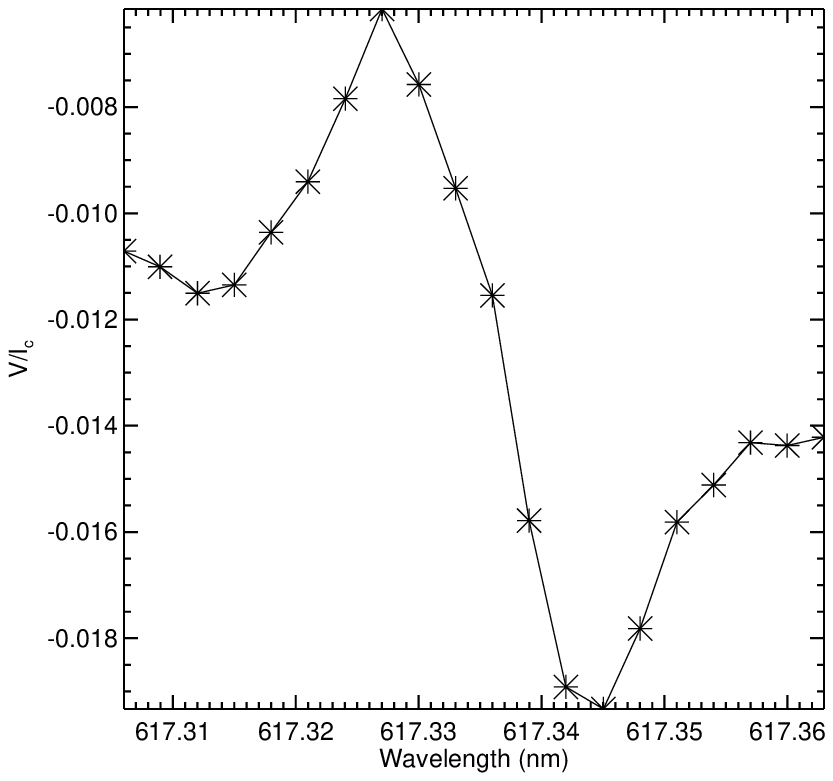}
             }
    \vspace{-0.35\textwidth}   % Shift close to the panel top 
    \centerline{\Large \bf     % Includes the labels (here needs the color package)
     \hspace{0.0 \textwidth} %\color{white}{}
     \hspace{0.415\textwidth}  %\color{white}{}
        \hfill}
    \vspace{0.31\textwidth}    % Shift back to the panel bottom 
             
\caption{Stokes I and V observations of the active region NOAA AR 12529 in the spectral line 6173 \AA{}. Top Row: Mean intensity (Stokes I) image (left) and 
its mean intensity profile (right). Bottom Row: Mean Stokes V image (left) and its mean profile (right).}
   %\label{fig_20}
   \end{figure}

   \subsection{Stokes V measurement in CaII 8542 \AA{} line}
In this section, we report the circular polarization measurements obtained in the chromospheric CaII 8542 \AA{} line. 
The measurements were carried out with the second pair of LCVRs specifically procured for this wavelength. 
The imager is tuned to the blue wing $-150$ m\AA{} from the line center. The LCVR is sequentially switched between 
voltages corresponding to the modulation voltages for the left and right circular polarizations. A pair of 100 images 
were obtained for this measurement with an exposure time of 120 ms for each image. In Figure 20, left image shows 
one of the selected I+V images from these observations, whereas the right panel shows the mean V image at the above wavelength point. A clear Stokes V signal 
is present in the difference image (Figure 20, right), the  seeing variations and the long exposure times produce artifacts, which will definitely reduce by 
the ongoing adaptive optics installation. In the above observations, the images were taken only at one wavelength position, but the filter can be tuned to a 
considerable part of the CaII 8542 \AA{} line profile, which will be done in the future observations. The linear polarization measurement is also planned for this 
line.

  \begin{figure}   
   \centerline{\hspace*{0.015\textwidth}
               \includegraphics[width=0.5\textwidth,clip=]{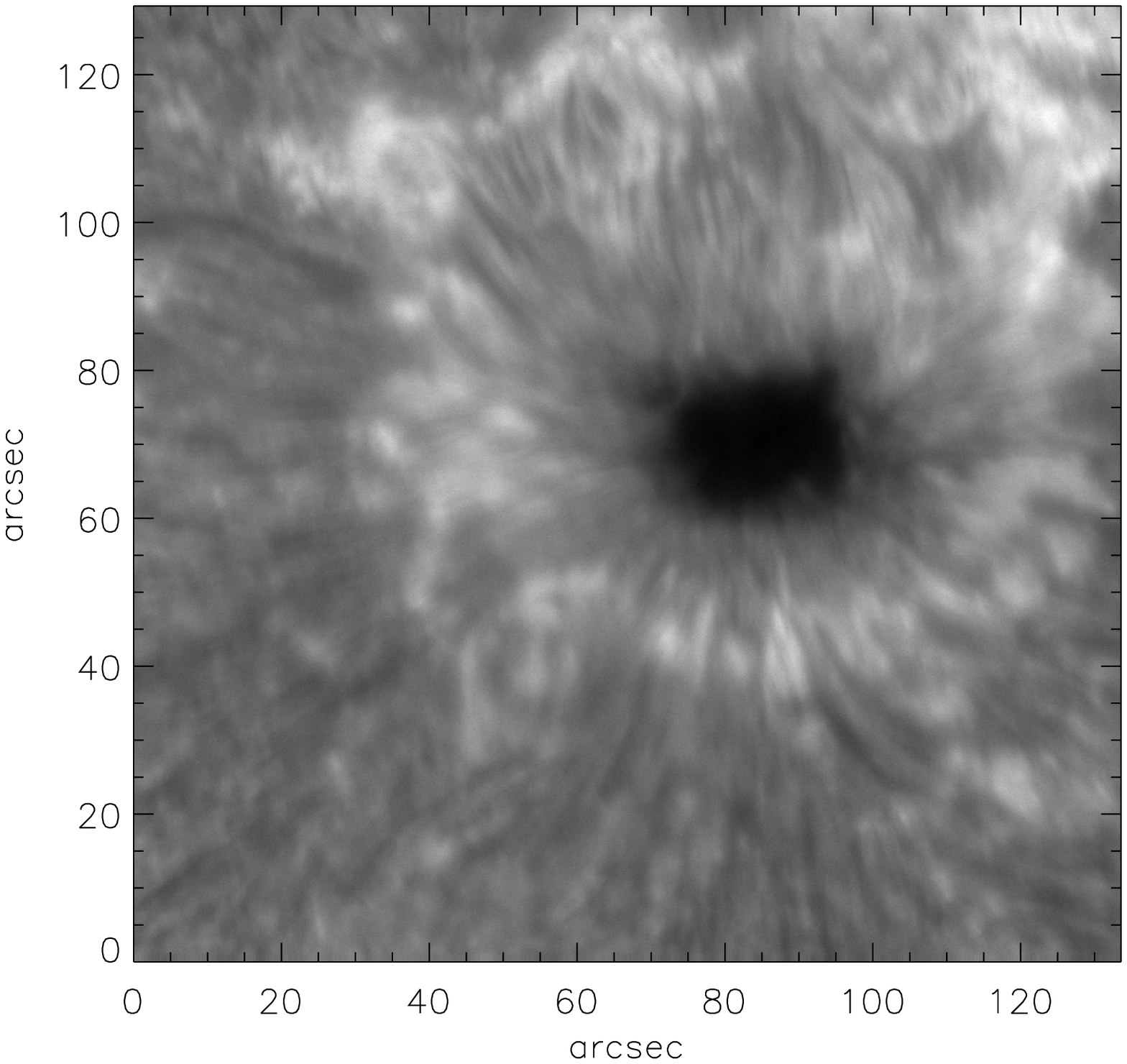}
               \hspace*{-0.03\textwidth}
               \includegraphics[width=0.5\textwidth,clip=]{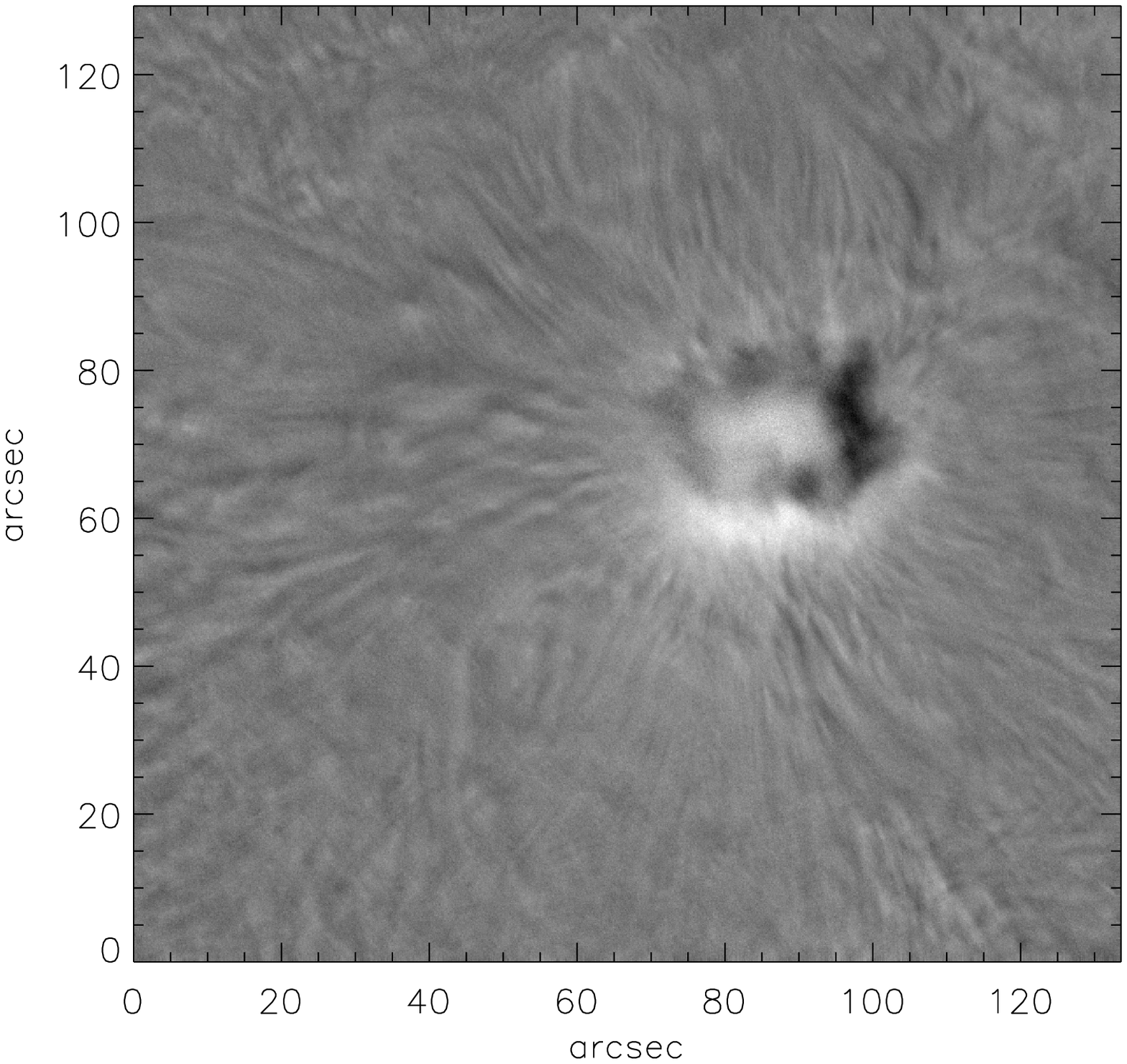}
              }
     \vspace{-0.35\textwidth}   % Shift close to the panel top 
     \centerline{\Large \bf     % Includes the labels (here needs the color 
                                %   package, see beginning of this file)
      \hspace{0.0 \textwidth}  %\color{white}{(a)}
      \hspace{0.5\textwidth}  %\color{white}{(b)}
         \hfill}
     \vspace{0.31\textwidth}    % Shift back to the panel bottom 
\caption{Chromospheric observations in LOS mode of the active region NOAA AR 12546 (S07, E19) observed on $18^{th}$ May, 2016 between 08:30 UT and 08:36 UT using MAST polarimeter.
Images of I+V (left) and mean Stokes parameter V (right) at a wavelength position $-150$ m\AA{} from the line center.}
   %\label{fig_21}
   \end{figure}

\section{Summary}
The imaging spectropolarimeter for MAST has been developed for obtaining the magnetic field information of the Sun in the spectral lines 6173 \AA{} and 8542 \AA{}, 
which are formed in photosphere and chromosphere, respectively. The spectropolarimeter includes an FP-based narrow-band filter, and a polarimeter consists of a pair 
of nematic LCVRs and a linear polarizer. In this paper, we have presented the characterization of the LCVRs, its retardance as a function of voltage and 
temperature.Response matrix of the polarimeter is obtained using an experimental setup. We have also discussed the implementation of four and 
six measurement schemes that are normally employed in obtaining the spectropolarimetric observations. 
Using the information obtained from the characterization of LCVRs, we have obtained preliminary observations in Fe I 6173 \AA{}. 
For the testing purpose, these observations are acquired by scanning the line profile of Fe I 6173 \AA{} at 27 wavelength positions with a sample of 15 m\AA{}. 
We plan to minimize the number of wavelength positions to $8-12$ to improve the cadence of the observations. As HMI also provides the similar observations, 
we compared the Stokes I and V observations from MAST with that of SDO/HMI. 
Qualitatively, both the observations are in good agreement with each other, considering the fact that MAST observations are seeing limited.

In order to obtain the vector magnetic fields of the active region, Stokes Q, and U along with Stokes I, and V are also obtained. However, we have not derived 
the magnetic fields from these observations as it requires information regarding instrument induced polarization. In this regard, it is important to note that 
MAST is a nine mirror system with two off-axis parabolic mirrors and 7 plane oblique mirrors. We have planned to obtain the instrument induced polarization both 
theoretically \citep{2015SPIE.9654E..08A, Sen1997} and experimentally \citep{Selbing2010}. In this paper, we have also presented the Stokes I and V observations of 
an active region in Ca II 8542 \AA{}, which is formed in the chromosphere.

\section{Acknowledgement}  We sincerely thank the referee for valuable comments, which helped us to improve the content in the manuscript. 
The HMI data used here are courtesy of NASA/SDO and the HMI science team. We thank the HMI science team for making available the processed data 
of $I-V$ (LCP) and $I+V$ (RCP) for our comparative study. We also acknowledge the work of Mukesh M Sardava of USO in the design and 
fabrication of the mount for LCVRs and polarizers. 
 
%   \bibliographystyle{spr-mp-sola}
%    \bibliography{paper_ref1}

% \bibliographystyle{spr-mp-sola}

\end{article} 
\end{document}